\documentclass[11pt,a4paper]{article}
\pdfoutput=1
\usepackage{jcappub}

\usepackage{comment}
\usepackage{amsmath}
\usepackage{amsfonts}
\usepackage{amssymb}
\usepackage{graphicx}
\usepackage{mathrsfs}
\usepackage{latexsym}
\usepackage{color}
\usepackage{slashed, cancel}
\usepackage{hyperref}
\usepackage{cleveref}
\usepackage{float}
\usepackage{tikz}
\usepackage{bbold}
\usepackage{soul} 
\usepackage{mathtools}
\usepackage{support-caption} 
\usepackage{subcaption}
\usepackage{enumitem}
\usepackage{multirow}
\usepackage{orcidlink}
\usepackage{journals}
\allowdisplaybreaks
\usepackage[version=3]{mhchem}
\hypersetup{urlcolor=blue, colorlinks=true} 

\usepackage{fancyhdr}
\numberwithin{equation}{section}

\definecolor{orange}{rgb}{1,0.4,0}
\definecolor{green}{rgb}{0,0.65,0}
\definecolor{rossos}{rgb}{0.8,0.2,0.3}
\definecolor{bluscuro}{rgb}{0.15, 0.2, .85}
\definecolor{bluchiaro}{cmyk}{1,.3,0.,0.1}
\hypersetup{colorlinks, citecolor=bluscuro, linkcolor=bluscuro, urlcolor=bluscuro}

\makeatother   
\newcommand{\PeV}{{\rm \,PeV}}
\newcommand{\GeV}{{\rm \,GeV}}
\newcommand{\TeV}{{\rm \,TeV}}
\newcommand{\MeV}{{\rm \,MeV}}

\newcommand{\cm}{{\rm \,cm}}
\newcommand{\fm}{{\rm \,fm}}
\newcommand{\km}{{\rm \,km}}
\newcommand{\s}{{\rm \,s}}

\newcommand{\K}{{\rm \,K}}
\newcommand{\kyr}{{\rm \,kyr}}
\newcommand{\Myr}{{\rm \,Myr}}
\newcommand{\Gyr}{{\rm \,Gyr}}
\newcommand{\yr}{{\rm \,yr}}
\newcommand{\yrs}{{\rm \,yrs}}

\newcommand{\Msun}{M_\odot}
\newcommand{\Mstar}{M_\star}
\newcommand{\Rstar}{R_\star}
\newcommand{\vstar}{v_\star}
\newcommand{\tstar}{t_\star}
\newcommand{\Tstar}{T_\star}
\newcommand{\Tstareq}{T_{\rm eq}}

\newcommand{\Teq}{T_{\rm eq}}

 \newcommand{\qomax}{q_0^{\rm MAX}}

\newcommand{\fMB}{f_{\rm MB}}
\newcommand{\fFD}{f_{\rm FD}}

\newcommand{\sigmathi}{\sigma^{th}_{i\chi}}
\newcommand{\kinFi}{\varepsilon_{F,i}}
\newcommand{\kinFn}{\varepsilon_{F,n}}
\newcommand{\tkin}{t_{\rm kin}}
\newcommand{\teq}{t_{\rm eq}}
\newcommand{\tth}{t_{\rm therm}}
\newcommand{\tthn}[1]{\tth^{(n=#1)}}

\newcommand{\mbeff}{m_i^{\rm eff}}
\newcommand{\mneff}{m_n^{\rm eff}}

\newcommand{\mcrit}{m_\chi^{\rm crit}}

\newcommand{\sigmav}{\langle \sigma_{\rm ann} v_\chi \rangle}

\newcommand{\erf}{{\rm \,Erf}}
\newcommand{\Msq}{|\overline{M}|^2}

\begin{document}

\hfill KCL-PH-TH/2023-71

\hfill FERMILAB-PUB-23-810-T

\title{Thermalization and Annihilation of Dark Matter in Neutron Stars}

\author[a]{Nicole F.\ Bell\,\orcidlink{0000-0002-5805-9828},}
\author[b]{Giorgio Busoni\,\orcidlink{0000-0002-8527-0768},}
\author[c,d]{Sandra Robles\,\orcidlink{0000-0002-6046-8217}}
\author[a]{and Michael Virgato\,\orcidlink{0000-0002-8396-0896}}
\affiliation[a]{ARC Centre of Excellence for Dark Matter Particle Physics, \\
School of Physics, The University of Melbourne, Victoria 3010, Australia}
\affiliation[b]{ARC Centre of Excellence for Dark Matter Particle Physics, \\
Research School of Physics,
The Australian National University, Canberra ACT 2601, Australia}
\affiliation[c]{Theoretical Particle Physics and Cosmology Group, Department of Physics, King’s College London, Strand, London, WC2R 2LS, UK}
\affiliation[d]{Particle Theory Department, Theory Division, Fermi National Accelerator Laboratory, Batavia, Illinois 60510, USA}
\emailAdd{n.bell@unimelb.edu.au}
\emailAdd{giorgio.busoni@anu.edu.au}
\emailAdd{srobles@fnal.gov}
\emailAdd{mvirgato@student.unimelb.edu.au}

\abstract{
The capture of dark matter, and its subsequent annihilation, can heat old, isolated neutron stars.  
In order for kinetic heating to be achieved, the captured dark matter must undergo sufficient scattering to deposit its kinetic energy in the star. We find that this energy deposit typically occurs quickly, for most of the relevant parameter space. In order for appreciable annihilation heating to also be achieved, the dark matter must reach a state of capture-annihilation equilibrium in the star. We show that this can be fulfilled for all types of dark matter - baryon interactions. This includes cases where the scattering or annihilation cross sections are momentum or velocity suppressed in the non-relativistic limit. Importantly, we find that capture-annihilation equilibrium, and hence maximal annihilation heating, can be achieved
without complete thermalization of the captured dark matter.
For scattering cross sections that saturate the capture rate, we find that capture-annihilation equilibrium is typically reached on a timescale of less than $1$~year for vector interactions and $10^4$~years for scalar interactions.}

\maketitle

\section{Introduction}

There has been much recent work on the capture of dark matter (DM) 
in neutron stars (NSs)~\cite{Goldman:1989nd,Kouvaris:2007ay,Kouvaris:2010vv,deLavallaz:2010wp,Kouvaris:2010jy, McDermott:2011jp,Kouvaris:2011fi,Guver:2012ba,Capela:2013yf,Bell:2013xk,  Bramante:2013nma,Bramante:2017xlb,Baryakhtar:2017dbj, Raj:2017wrv, Bell:2018pkk,  Garani:2018kkd,Bell:2019pyc, Garani:2019fpa, Acevedo:2019agu, Joglekar:2019vzy, Joglekar:2020liw, Bell:2020jou, Dasgupta:2020dik, Bell:2020lmm, Bell:2020obw,Leane:2021ihh,Anzuini:2021lnv,Bose:2021yhz,Bramante:2021dyx, Tinyakov:2021lnt, DeRocco:2022rze, Fujiwara:2022uiq,Hamaguchi:2022wpz,Coffey:2022eav, Chatterjee:2022dhp, Nguyen:2022zwb, Alvarez:2023fjj,Bhattacharya:2023stq,Bramante:2023djs}, as a sensitive probe of dark matter interactions with ordinary matter. 
This can potentially be used to test dark matter interactions in a way that is highly complementary to experiments on Earth, especially since DM is accelerated to relativistic speeds during infall to a NS.
In some  cases, neutron star techniques could potentially probe interactions that would be difficult or impossible to ever observe in dark matter direct detection experiments. This includes dark matter that is too light to leave a detectable signal in nuclear-recoil experiments, or interactions for which the non-relativistic scattering cross section is momentum suppressed.

It was recently pointed out that old, isolated, NSs in the Solar neighborhood could be heated by DM capture~\cite{Baryakhtar:2017dbj}, leading to a temperature increase of $\sim 2000\K$. 
 At ages greater than $\sim 10\Myr$, isolated NSs are expected to cool to temperatures below this, provided they are not reheated by accretion of standard matter or by internal heating mechanisms~\cite{Gonzalez:2010ta}.
As a result, the observation of a local NS with a temperature $\mathcal{O}(1000\K)$ could provide stringent constraints on DM interactions. Importantly,
NS temperatures in this range would result in near-infrared emission, potentially detectable by future telescopes.

There are two distinct contributions to this heating: 
\begin{enumerate}
    \item Kinetic heating, where the DM kinetic energy is deposited in the NS medium. 
    \item Annihilation heating, where the DM rest mass energy is deposited through the annihilation of DM to particles that are trapped in the star.
\end{enumerate}

The kinetic heating occurs as follows: The initial scattering interaction, which leads to gravitational capture of the DM particle, transfers only a small portion of the DM kinetic energy to the star.  The rest of the kinetic energy is transferred through subsequent scattering interactions of the gravitationally bound DM until, eventually, the DM thermalizes with constituents in the centre of the star.  In general, a large number of collisions are typically required for full thermalization.  Moreover, if the scattering cross section is momentum suppressed in the non-relativistic limit, the time between collisions, and hence the time required to achieve full thermalization, can become very large, larger than the age of the Universe.  Importantly, we shall see that even in cases where the full thermalization process is slow, {\it the majority of the kinetic energy is deposited very quickly.}

DM annihilation occurs in a region very close to the centre of the star, where the thermalized DM accumulates.  The annihilation rate will increase as the DM abundance in the star grows until, eventually, a state of  equilibrium between capture and annihilation is reached.  When this occurs, the annihilation heating is maximized.  The complete thermalization of the DM will result in a smaller, denser sphere of thermalized DM in the centre of the star, assisting in annihilation.  However, we shall find that {\it capture-annihilation equilibrium, and hence maximal annihilation heating, can be achieved \it without full thermalization}.

The thermalization process was previously examined in refs.~\cite{Bertoni:2013bsa, Garani:2020wge} for a subset of interaction types. These previous studies did not consider the importance of annihilation heating from partially thermalized DM. In this paper, we present a detailed calculation of the timescales required for thermalization and for capture-annihilation equilibrium, utilising our improved treatment of DM capture in NSs~\cite{Bell:2020jou,Bell:2020lmm,Bell:2020obw,Anzuini:2021lnv}. We perform these calculations for a full set of DM-nucleon interaction types, parameterized by a set of Effective Field Theory (EFT) operators for fermionic DM.  This includes operators for which either the scattering cross section or the annihilation cross section is suppressed in the non-relativistic limit.  By properly accounting for the annihilation of partially thermalized DM, we show that full kinetic plus annihilation heating can be achieved for most of the interesting parameter space, on a short timescale.

This paper is organized as follows: We briefly review NS structure and composition in  Section~\ref{sec:ns}, and outline the calculation of the NS temperature due to DM energy deposition in Section~\ref{sec:temperature}.
Section~\ref{sec:capintrates} reviews the DM capture and interaction rates, while Section~\ref{sec:thermalization} examines the thermalization process. The annihilation of captured DM is discussed in Section~\ref{sec:CAEquilibrium}, where we explain how to modify the usual capture-annihilation criterion to account for the annihilation of partially thermalized DM. The timescales for kinetic and annihilation heating are discussed in Section~\ref{sec:results} and concluding remarks given in Section~\ref{sec:conclusions}.

\section{Neutron Star Composition}
\label{sec:ns}

Massive stars, with masses above $\sim 8-10\Msun$ and below $\sim 20\Msun$, end their life in core-collapse supernova explosions, leaving behind the densest known compact stellar object, a neutron star. The strongly degenerate matter found in a NS provides a unique means to test fundamental interactions and to look for new physics. Despite recent theoretical and observational breakthroughs, many uncertainties remain regarding NS composition and their equation of state (EoS).

Several phase transitions are expected to be found towards the interior of a NS. Below  a thin atmosphere, the stellar interior can be broadly divided into a crust and a core. The outer layers of the former insulate the surface from the hot core. The core accounts for $\sim99\%$ of the stellar mass and is composed mostly of 
strongly degenerate neutrons. However, as the density increases towards the centre of these stellar remnants, beta equilibrium allows for the existence of protons, electrons and muons; the so-called $npe\mu$ matter.
In addition, the extremely dense inner core of massive NSs may harbor exotic species, such as hyperons, which also appear under beta equilibrium. 

For a consistent estimation of the DM capture rate, thermalization times and annihilation rate,  an equation of state (EoS) that relates pressure and density is needed in order to solve 
the NS structure equations.  Of the several EoSs of dense matter available in the literature \cite{Akmal:1998cf,RikovskaStone:2006ta,Goriely:2010bm,Goriely:2013,Kojo:2014rca,Baym:2017whm,Annala:2019eax},  we have chosen the quark-meson coupling (QMC) model~\cite{Guichon:2018uew,Motta:2019tjc}  as in refs.~\cite{Bell:2020obw,Anzuini:2021lnv}. 
This relativistic EoS is consistent with current constraints and permits the existence of $\sim2\Msun$ NSs that contain hyperonic matter~\cite{RikovskaStone:2006ta}. 
We also assume a non-rotating, non-magnetized, spherically symmetric NS. We then solve the coupled system of Tolman-Oppenheimer-Volkoff (TOV) equations \cite{Tolman:1939jz,Oppenheimer:1939ne} and QMC EoS for given central baryon number densities $n_{\cal B}^c$ (see Table~\ref{tab:eos}), from the NS centre to the surface. In this way, the radius, $\Rstar$, and the gravitational mass, $\Mstar$, of the star are determined, as well as radial profiles for the number density and effective mass $\mbeff$ of each particle species, and general relativity corrections embedded in $B(r)$ (the coefficient that appears in the time part of the Schwarzschild metric). For an example of these radial profiles see Fig.~1 of refs.~\cite{Bell:2020jou,Anzuini:2021lnv} and  Fig.~2 of ref.~\cite{Bell:2020lmm}. 

In Table~\ref{tab:eos}, we show the relevant NS quantities needed for our calculations, for three of the four NS configurations obtained in ref.~\cite{Anzuini:2021lnv}. These three benchmark stars will be used to illustrate the dependence of our results on the compactness of the NS.

\begin{table}[tb]
\centering
\begin{tabular}{|l|c|c|c|c|c|c|c|}
\hline
\bf EoS & $n_{\cal B}^c$ $[\fm^{-3}]$ & $\Mstar$ $[\Msun]$ & $\Rstar$ [km] & $\mneff(0)$ [GeV] & $\kinFn(0)$ [GeV] & $B(0)$ & $B(\Rstar)$\\ \hline
\bf QMC-1 & 0.325 & 1.000 & 13.044 & 0.592 &  0.127 & 0.580 & 0.772 \\
\bf QMC-2 & 0.447 & 1.500 & 12.847 & 0.540 & 0.162 & 0.408 & 0.653 \\
\bf QMC-3 & 0.872 & 1.900 & 12.109 & 0.463 &  0.180 & 0.235 & 0.535  \\
\hline
\end{tabular} 
\caption{Properties of the benchmark NS configurations as determined in refs.~\cite{Bell:2020obw,Anzuini:2021lnv}, using the QMC equation of state and varying the central baryon number density $n_{\cal B}^c$.  From left to right, these quantities are NS mass; radius; neutron effective mass $\mneff(0)$ and Fermi energy $\kinFn(0)$ at the NS centre; and $B(0)$ and  $B(\Rstar)$, which are related to the escape velocity.}
\label{tab:eos}
\end{table} 

\section{Neutron Star Temperature from Dark Matter  Heating}
\label{sec:temperature}
We now discuss the potential NS heating that can be achieved by DM scattering and annihilation. 
We assume a nearby NS, located in the Solar neighborhood, and thus take $\rho_\chi=0.4\GeV\cm^{-3}$, $\vstar=230\km\s^{-1}$ and $v_d=270\km\s^{-1}$ as the DM density, NS velocity, and DM dispersion velocity respectively. 
DM deposits energy into the NS via two mechanisms: (i) kinetic heating due to scattering with the constituents of the NS and (ii) annihilation of DM to SM particles that do not escape the star. 
We define the DM contribution to the temperature, measured by an observer far from the star, as $T^\infty_{\chi}=\sqrt{B(\Rstar)} \, T_\chi$, where, $B(r)$ is the time component of the Schwarzchild metric.  
Assuming black body radiation, this temperature will be given by 
\begin{equation}
        T^\infty_{\chi}  = \left[\frac{B^2(\Rstar)}{4\pi\sigma_\text{SB} R_\star^2} \dot{E}_{\chi}  \right]^{1/4},
\label{eq:Tkin} 
\end{equation}
where $\dot E_{\chi}= \dot E_{\chi, \rm kin} + \dot E_{\chi, \rm ann}$ is the rate of energy deposition and we have assumed the absence of any other source of heating.

The DM kinetic energy is deposited at the rate
\begin{align}
    \dot E_{\chi, \rm kin} & \simeq m_\chi \left( \frac{1}{\sqrt{B(0)}} - 1\right)C_{\rm geom}f, \label{eq:kinenergy}
    \end{align}
    where $C_{\rm geom}$ is the maximum DM capture rate (defined later in Eq.~\ref{eq:capturegeom})    and $f$ quantifies how efficiently DM is captured, 
    \begin{align}
    f & \simeq \min\left[ 1,  \frac{ \sum_i C_i}{C_{\rm geom}} \right],
\end{align}
where we sum over the capture rates $C_i$ for scattering on all baryonic species $i$ in the star. 
Note that we have used $B(0)$ in Eq.~\ref{eq:kinenergy}, instead of $B(\Rstar)$, which was previously used in the literature. This is because gravitational potential energy is converted to kinetic energy as the DM falls deeper into the NS. Therefore, the total energy the DM can deposit is equal to the kinetic energy it gains when moving from infinity to the centre of the star. 
If this were the only source of heating, the observed temperature would be  $T^\infty_{\chi,\rm kin}  \sim 1870\K\;f^{1/4}$ for the QMC-2 ($1.5\Msun$) benchmark NS. For the $1\Msun$ and $1.9\Msun$ NSs, we find 
$\sim 1510\K\; f^{1/4}$ and $\sim2240\K\; f^{1/4}$, respectively.

Annihilation of DM in the centre of the NS causes further heating. The annihilation rate $\Gamma_{\rm ann}$, and hence the annihilation heating, is maximized when capture-annihilation equilibrium has been achieved. In this limit, the DM annihilation rate is given by $\Gamma_{\rm ann} = C/2$. Then, the rate at which DM deposits all of its energy, both kinetic and rest-mass, can be expressed as 
\begin{equation}
    \dot E_{\chi, \rm kin + ann} = \dot E_{\chi, \rm kin} + 2 \Gamma_{\rm ann} m_\chi \simeq \frac{m_\chi}{\sqrt{B(0)}} C_{\rm geom}f. \label{eq:massheating}
\end{equation} 
This rate implies a temperature of $T^\infty_{\chi,\rm kin+ann}  \sim  2410\K\;f^{1/4}$ for the $1.5\Msun$ NS. For the lightest NS considered ($1\Msun$) this value decreases to $\sim  2160\K$, while for the heaviest NS ($1.9\Msun$), this temperature reaches $\sim  2640\K\;f^{1/4}$. Therefore, annihilation heating contributes an additional $\sim 400-650\K$ to the NS temperature compared to kinetic heating alone, depending on the NS configuration.

\section{Scattering and Capture Rates}
\label{sec:capintrates}

In this paper, we restrict our attention to DM scattering with baryons, the most abundant particle species in NSs.\footnote{DM-lepton scattering can be treated in an analogous way. Note that, in general, the DM-lepton couplings will be independent of the DM-quark/hadron couplings. 
} To compute the DM scattering and capture rates in neutron stars, we must account for a number of effects including relativistic kinematics, general relativistic corrections, the motion of the target particles in the NS, and Pauli blocking (relevant to the scattering of light DM)~\cite{Bell:2020jou,Bell:2020lmm}.  In addition, the extreme conditions in the NS interior give rise to two additional important effects: (i) baryons experience strong interactions, which result in an effective mass different from their in vacuum value, and (ii) the momentum transfer in each collision is large enough to resolve the internal structure of the nucleon~\cite{Bell:2020obw,Anzuini:2021lnv}.\footnote{In addition, DM-phonon scattering in the NS core will play a role in the dynamics at very low momentum transfer if baryons are in a superfluid state~\citep{Bertoni:2013bsa}.}
A detailed account of the scattering and capture rate calculations can be found in refs.~\cite{Bell:2020jou,Bell:2020lmm,Bell:2020obw,Anzuini:2021lnv}. Here, we briefly outline the aspects of the scattering calculation relevant to this paper.

\subsection{Scattering Rate}
\label{sec:intratetext}

The DM-baryon scattering rate is a key ingredient in both the capture and thermalization processes. 
It is used in constructing the DM energy loss probability distribution function, which, in turn, determines the probability that the DM is captured~\cite{Bell:2020jou}, and the average energy lost in each collision during the thermalization process.

The most general expression for the DM down-scattering rate, expressed in terms of the DM-target response function $S(q_0,q)$~\cite{Bertoni:2013bsa,Bell:2020jou}, is 
\begin{equation}
\Gamma^-(K_\chi) =   \int \frac{d^3k'}{(2\pi)^3} \frac{1}{(2K_\chi)(2K'_\chi)(2m_i)(2m_i)} \Theta(E'_\chi-m_\chi)\Theta(q_0)S(q_0,q), \label{eq:intratedeftext}
\end{equation} 
where
\begin{align}
S(q_0,q)  =   2 & \int \frac{d^3p}{(2\pi)^3}\int \frac{d^3p'}{(2\pi)^3} \frac{m_i^2}{E_i E'_i}|\overline{M}(s,t,m_i)|^2 
(2\pi)^4\delta^4\left(k_\mu+p_\mu-k'_\mu-p'_\mu\right)\nonumber\\
 &\times\fFD(E_i)(1-\fFD(E'_i)) \Theta(E_i-m_i)\Theta(E'_i-m_i),
 \label{eq:responsefunc}
\end{align}
$k^\mu=(K_\chi$,$\vec{k})$ and $k'^{\mu}=(K'_\chi,\vec{k'})$ are the DM initial and final momenta,  $p^\mu=(E_i,\vec{p})$ and $p'^{\mu}=(E'_i,\vec{p'})$ are the target particle initial and final momenta, $q_0=E'_i-E_i$ is the DM energy loss, $\vec{q}=\vec{p}-\vec{p'}$ is the three-momentum exchanged in the scattering, $m_i$  is the mass of the target, $|\overline{M}(s,t,m_i)|^2$ is the  spin-averaged squared matrix element, and $\fFD$ is the Fermi Dirac distribution. Note that the purpose of $\Theta(q_0)$ is to select only the down-scattering interactions.

When the energy transfer is small compared to the Fermi energy, Pauli blocking strongly suppresses the scattering rate.  Since the targets are in a highly degenerate Fermi plasma, we can take the target energy levels to be either full or empty in the zero temperature limit. Therefore, the interaction rate depends on the number of targets with energy $E_i$ in the initial state, and on the number of empty final states with energy $E_i+q_0$, where $q_0$ is the dark matter energy loss. As a result, the interaction rate necessarily vanishes in the limit that  $q_0\rightarrow 0$.  More generally,  Pauli blocking suppresses the differential interaction rate when $E_i+q_0\le \kinFi$.

To calculate the squared matrix elements, $|\overline{M}|^2$, we take the DM-quark couplings to be described by the four-fermion effective field theory (EFT) operators listed in Table~\ref{tab:operatorshe}, where the strength of the coupling is parameterized by a cutoff scale, $\Lambda_q$. For each of these operators, the corresponding $|\overline{M}|^2$ is expressed in terms of the Mandelstam variables $s$ and $t$, and the DM-target mass ratio 
\begin{equation}
    \mu=\frac{m_\chi}{m_i}.
\end{equation}
In the following sections, we shall derive approximations for interaction rates and timescales that depend on the form of the matrix element. We therefore define the parameter $n$ to denote the $t$-dependence of  the squared matrix element as 
\begin{equation}
\Msq \propto t^n, 
\label{eq:ndef}
\end{equation}
with $n=0,1,2$. 

For DM interactions with baryons, the squared matrix elements contain hadronic coefficients, $c_i^I(t)$, which depend on the transferred momentum $t$. In the case of scalar and pseudoscalar interactions, these coefficients also depend on the baryon mass $m_i$. 
In general, these coefficients can be defined as a function of their values at zero momentum transfer as $c_i^I(0)$~\cite{Thomas:2001kw}
\begin{eqnarray}
c_i^I(t) &= c_i^I(0) F(t),\quad I\in\{S,P,V,A,T\},\label{eq:tdep}
\end{eqnarray}
where S, P, V, A and T denote scalar, pseudoscalar, vector, axial and tensor interactions, respectively. The coefficients $c_i^I(0)$  are given in appendix~A of ref.~\cite{Anzuini:2021lnv}, while $F(t)$ is the square of the dipole form-factor,
\begin{equation}
    F(t) = \frac{1}{(1-t/Q_0^2)^4}. 
    \label{eq:formfactor}
\end{equation}
Note that the energy scale $Q_0$ depends on the specific hadronic form factor. As in refs.~\cite{Bell:2020obw,Anzuini:2021lnv}, we conservatively assume $Q_0=1\GeV$ for all operators and baryonic target species. 
The effect of the strong interactions between the baryons is incorporated in the proper calculation of the Fermi energies of the baryonic species, $\kinFi$, and in the use of baryon effective masses for the target masses. These depends on the microphysics embedded in the EoS~\cite{Bell:2020obw,Anzuini:2021lnv}.

For the collisions that result in DM capture, analytic scattering rate expressions can be obtained~\cite{Bell:2020jou, Bell:2020lmm} because the DM kinetic energy is always significantly higher than the NS temperature and hence the zero temperature $\Tstar\rightarrow0$ approximation holds. 
During the thermalization process, however, the DM kinetic energy and NS temperature become comparable, and so finite temperature effects become important. 
It is numerically intensive to calculate the interaction rate directly from Eq.~\ref{eq:intratedeftext} for these low temperatures, hence we keep only the lowest order thermal corrections.
This amounts to a modification of the differential interaction rate, i.e. the $q_0$ integrand of Eq.~\ref{eq:intratedeftext}, such that
\begin{equation}
    \frac{d }{d q_0}\Gamma^-(K_\chi, T_\star) = \frac{1}{1 - \exp(-q_0/T_\star)} \frac{d }{d q_0}\Gamma^-(K_\chi, T_\star = 0).
    \label{eq:diffGammaFiniteTemp}
\end{equation}
As a result, we can no longer obtain fully analytic expressions for the interaction rate. Therefore, all the results we present below have been calculated numerically.

\begin{table}
\centering
\setlength{\tabcolsep}{0.25em}   
\begin{tabular}{ | c | c | c | c | c | c |}
  \hline                        
  \multirow{2}{*}{Name} & \multirow{2}{*}{Operator} & \multirow{2}{*}{$g_q$} & \multirow{2}{*}{$g_i^2$} & \multirow{2}{*}{$|\overline{M}(s,t,m_i)|^2$}   & Dominant \\[-0.3em]
  & & & & & term $\tth$\\   \hline
  D1 & $\bar\chi  \chi\;\bar q  q $ & $\frac{y_q}{\Lambda_q^2}$ & $\frac{c_i^S(t)}{\Lambda_q^4}$ & $ g_i^2(t)\frac{\left(4 m_{\chi }^2-t\right) \left(4 m_{\chi }^2-\mu ^2   t\right)}{\mu ^2}$ & $t^0$ \\  \hline
  D2 & $\bar\chi \gamma^5 \chi\;\bar q q $ & $i\frac{y_q}{\Lambda_q^2}$ & $\frac{c_i^S(t)}{\Lambda_q^4} $ & $g_i^2(t)\frac{t \left(\mu ^2 t-4 m_{\chi }^2\right)}{\mu ^2}$ & $t^1$ \\  \hline
  D3 & $\bar\chi \chi\;\bar q \gamma^5  q $&  $i\frac{y_q}{\Lambda_q^2}$ & $\frac{c_i^P(t) }{\Lambda_q^4}$ &  $g_i^2(t) t \left(t-4 m_{\chi }^2\right)$ & $t^1$\\  \hline
  D4 & $\bar\chi \gamma^5 \chi\; \bar q \gamma^5 q $ & $\frac{y_q}{\Lambda_q^2}$ & $\frac{c_i^P(t)}{\Lambda_q^4}$ & $g_i^2(t) t^2$  & $t^2$ \\  \hline
  D5 & $\bar \chi \gamma_\mu \chi\; \bar q \gamma^\mu q$ & $\frac{1}{\Lambda_q^2}$ & $\frac{c_i^V(t)}{\Lambda_q^4}$ &  $2 g_i^2(t) \frac{2 \left(\mu ^2+1\right)^2 m_{\chi }^4-4 \left(\mu ^2+1\right) \mu ^2 s m_{\chi }^2+\mu ^4 \left(2 s^2+2 s t+t^2\right)}{\mu^4}$ & $t^0$ \\  \hline
  D6 & $\bar\chi \gamma_\mu \gamma^5 \chi\; \bar  q \gamma^\mu q $ & $\frac{1}{\Lambda_q^2}$ & $\frac{c_i^V(t)}{\Lambda_q^4}$ & $2  g_i^2(t)\frac{2 \left(\mu ^2-1\right)^2 m_{\chi }^4-4 \mu ^2 m_{\chi }^2 \left(\mu ^2 s+s+\mu ^2 t\right)+\mu ^4 \left(2 s^2+2 s   t+t^2\right)}{\mu^4}$ & $t^0$ \\  \hline
  D7 & $\bar \chi \gamma_\mu  \chi\; \bar q \gamma^\mu\gamma^5  q$ & $\frac{1}{\Lambda_q^2}$ & $\frac{c_i^A(t)}{\Lambda_q^4}$ &  $2  g_i^2(t) \frac{2 \left(\mu ^2-1\right)^2 m_{\chi }^4-4 \mu ^2 m_{\chi }^2 \left(\mu ^2 s+s+t\right)+\mu ^4 \left(2 s^2+2 s t+t^2\right)}{\mu^4}$ & $t^0$ \\  \hline
  D8 & $\bar \chi \gamma_\mu \gamma^5 \chi\; \bar q \gamma^\mu \gamma^5 q $ & $\frac{1}{\Lambda_q^2}$ & $\frac{c_i^A(t)}{\Lambda_q^4}$ &  $2  g_i^2(t) \frac{2 \left(\mu ^4+10 \mu ^2+1\right) m_{\chi }^4-4 \left(\mu ^2+1\right) \mu ^2  m_{\chi }^2 (s+t)+\mu ^4 \left(2 s^2+2 s t+t^2\right)}{\mu ^4}$ & $t^0$\\  \hline
  D9 & $\bar \chi \sigma_{\mu\nu} \chi\; \bar q \sigma^{\mu\nu} q $ & $\frac{1}{\Lambda_q^2}$ & $\frac{c_i^T(t)}{\Lambda_q^4}$ & $8  g_i^2(t) \frac{4 \left(\mu ^4+4 \mu ^2+1\right) m_{\chi }^4-2 \left(\mu ^2+1\right) \mu ^2 m_{\chi  }^2 (4 s+t)+\mu ^4 (2 s+t)^2}{\mu ^4}$ & $t^0$ \\  \hline
 D10 & $\bar \chi \sigma_{\mu\nu} \gamma^5\chi\; \bar q \sigma^{\mu\nu} q $ & $\frac{i}{\Lambda_q^2}$ & $\frac{c_i^T(t) }{\Lambda_q^4}$ &  $8  g_i^2(t)\frac{4 \left(\mu ^2-1\right)^2 m_{\chi }^4-2 \left(\mu ^2+1\right) \mu ^2 m_{\chi }^2 (4 s+t)+\mu ^4 (2 s+t)^2}{\mu^4}$ & $t^0$ \\  \hline
\end{tabular}
\caption{Dimension 6 EFT operators~\cite{Goodman:2010ku} for the coupling of Dirac DM to quarks (column 2), together with the squared matrix elements for DM-hadron scattering (column 5), where $s$ and $t$ are Mandelstam variables, $\mu=m_\chi/m_i$, and $m_i$ is the hadron mass.
The quark-level coefficients $g_q$ are expressed in terms of the Yukawa couplings, $y_q$, and the cutoff scale, $\Lambda_q$. The squared hadron-level coefficients, $g_i^2$, depend on the momentum-dependent couplings $c_i^I(t)$, as in Eq.~\ref{eq:tdep}. 
The final column indicates the dominant term in the squared matrix element, in the low momentum-transfer limit relevant for the late stage of the thermalization process. 
\label{tab:operatorshe} }
\end{table}

\subsection{Capture Rate}
\label{sec:caprate}

For DM-baryon cross sections much smaller than a threshold value, which we denote as $\sigmathi$,  a neutron star can be considered as an optically thin medium. In this regime, DM passes through the star and capture can occur anywhere in the NS interior. The capture rate can then be calculated using the following expression
\begin{equation}
C =  \frac{4\pi\rho_\chi}{m_\chi} 
\int_0^\infty du_\chi \frac{\fMB(u_\chi)}{u_\chi} 
\int_0^{\Rstar}  r^2 \frac{\sqrt{1-B(r)}}{B(r)} \Omega^{-}(r)  \, dr, \label{eq:capturefinalM2text} 
\end{equation}
where $\rho_\chi$ is the local DM density, $u_\chi$ is the DM velocity far away from the NS, 
and $\fMB$ is the distribution of relative velocities
between NS targets and DM particles far away from the NS, which we assume to be of Maxwell-Boltzmann form.  
We define $\Omega^{-}(r) $ to be the interaction rate of
Eq.~\ref{eq:intratedeftext} in the special case where the initial DM kinetic energy is fixed to $K_\chi = m_\chi(1/\sqrt{B(r)} - 1)$, as relevant for a DM particle infalling to the NS. 
This can be expressed as
\begin{eqnarray}
    \Omega^{-}(r) &=& \frac{1}{32\pi^3}\int dt dE_i ds  \frac{s|\overline{M}(s,t,\mbeff)|^2}{s^2-[(\mbeff)^2-m_\chi^2]^2} \frac{E_i}{m_\chi} \sqrt{\frac{B(r)}{1-B(r)}} \nonumber \\
&&\times\frac{\fFD(E_i,r)(1-\fFD(E_i^{'},r))}{\sqrt{[s-(\mbeff)^2-m_\chi^2]^2-4(\mbeff)^2m_\chi^2}},
\label{eq:intrate}
\end{eqnarray}
where the integration intervals for are given in refs.~\cite{Bell:2020jou,Anzuini:2021lnv}, together with further details about the capture rate calculations.

If the DM-target cross section reaches or surpasses the threshold value $\sigmathi$, the NS optical depth $\tau_\chi$ can no longer be taken as close to zero. We must then introduce an optical factor, $\eta(r)=\exp[-\tau_\chi(r)]$, in Eq.~\ref{eq:capturefinalM2text} to suppress the capture rate; see ref.~\cite{Bell:2020jou} for further details. 
In the optically thick limit, the NS is seen as a rigid sphere by the DM particles, and all of the capture occurs on the surface of the star. This is the so-called geometric limit. In this case, the capture rate is given by~\cite{Bell:2018pkk}
\begin{equation}
C_\textrm{geom} =  \frac{\pi R_\star^2[1-B(R_\star)]}{v_\star B(R_\star)} \frac{\rho_\chi}{m_\chi} \erf\left(\sqrt{\frac{3}{2}}\frac{v_\star}{v_d}\right).
\label{eq:capturegeom}    
\end{equation}

\section{Thermalization}
\label{sec:thermalization}

After becoming gravitationally bound to the NS, the DM particles continue to scatter with NS targets, losing energy in each collision until reaching thermal equilibrium at the centre of the star. 
We outline the calculation of the thermalization time in terms of the average DM energy lost in a single collision, and use first-order approximations to derive scaling relations that allow us to understand the qualitative features of our numerical results.

\subsection{Average DM energy loss}
\label{sec:energyloss}

\begin{figure}
\centering
\includegraphics[width=0.49\textwidth]{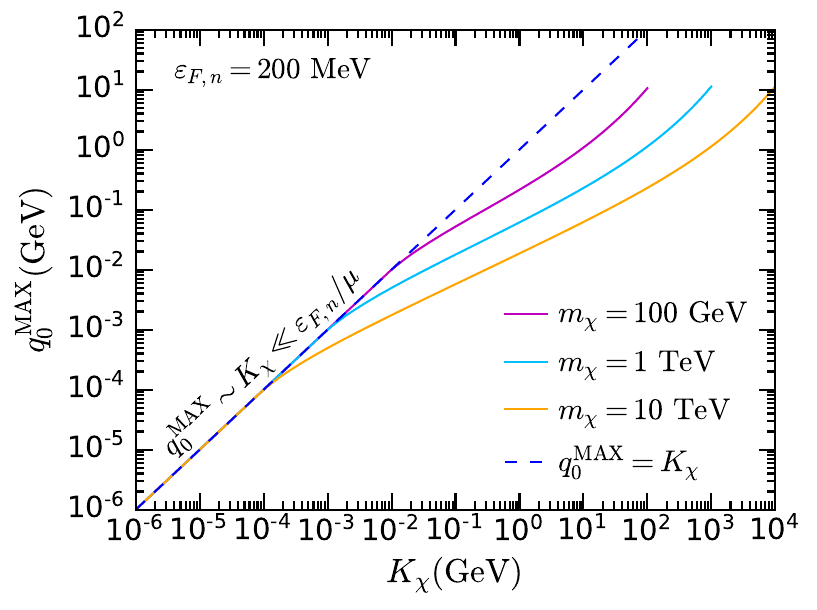}
\caption{Maximum energy loss per collision with neutron targets, as a function of the DM kinetic energy. We have assumed $\kinFn=200\MeV$.} 
\label{fig:q0max}
\end{figure}

The average energy a DM particle loses per collision can be calculated by weighting the DM energy loss, $q_0$, with differential interaction rate. We thus obtain
\begin{equation}
\langle \Delta K_\chi\rangle = \frac{1}{\Gamma^{-}} \int_{0}^{\qomax} dq_0 q_0  \frac{d\Gamma^-}{dq_0}, 
\end{equation}
where $\qomax$ is the maximum energy lost in a single scatter. 
Figure~\ref{fig:q0max} shows $\qomax$ as a function of the DM kinetic energy, $K_\chi=E_\chi-m_\chi$, for DM-neutron collisions. We see that heavier DM particles lose a smaller fraction of their kinetic energy per collision than lighter DM\footnote{In this figure, the additional suppression introduced by the momentum-dependent hadronic matrix elements is neglected.}.  
Nevertheless, as $K_\chi$ approaches the Pauli blocked region, $K_\chi \ll m_i \kinFn/m_\chi$ (dashed blue line), the maximum energy loss per collision becomes independent of the DM mass.

For the initial collision that results in capture, Pauli blocking represents, at most, a sub-leading correction to the capture rate for DM masses above the Fermi energy of the targets. 
Following capture, however, the DM energy will continue to decrease as a result of subsequent scattering, eventually reaching kinetic energies where Pauli blocking is an important effect. Consequently, Pauli blocking will strongly impact the rate at which dark matter is thermalized, for a wide DM mass range that extends well above $\kinFi$.

It is useful to define a critical DM mass, above which Pauli blocking is never in effect throughout the entire thermalization process. We do this by analysing the regions of the interaction rate phase space that are suppressed by Pauli blocking, arriving at
\begin{equation}
     m_\chi \gtrsim \frac{2\kinFi(2 m_i + \kinFi)}{K_\chi} =m_\chi^{\rm crit}. 
    \label{eq:mucrit}
\end{equation}
For neutron targets with $\kinFn=200\MeV$, and assuming an equilibrium temperature of $10^3\K$, i.e. $K_\chi \gtrsim 10^3 \K$, we find (neglecting the nucleon form factors) $m_\chi^{\rm crit} \sim 9.65\times 10^9\GeV$.\footnote{Note that at low energies, where $K_\chi \ll m_\chi$, and DM masses  $m_\chi\lesssim m_i \frac{\kinFi}{K_\chi}$,  the maximum DM energy loss in a single scattering is $\qomax\sim K_\chi\ll\kinFi$.} 
Pauli blocking will then suppress at least some part of the thermalization process for all DM masses below this value.

In either regime, we can obtain first-order approximations for the average fraction of energy that a DM particle loses in a single collision by making use of the zero temperature approximation, a constant target mass, and nucleonic form factors at zero momentum transfer, i.e., $F(t)\sim1$. 
First, we consider the regime in which Pauli blocking is negligible, $m_\chi \gtrsim m_\chi^{\rm crit}$. For a constant cross section (i.e. $\Msq \propto t^0$) we find 
\begin{equation}
\langle \Delta K_\chi^{(n=0)} \rangle 
\sim 2 \sqrt{\frac{\kinFi}{\mu}} K_\chi^{1/2} \ll K_\chi, 
\label{eq:aveElossn0largem}
\end{equation}
at first order in $m_\chi^{\rm crit}/m_\chi$. (See appendix \ref{sec:pauliblockingle} for the corresponding approximation of the interaction rate, Eq.~\ref{eq:intraten0largem}.) 
For cross sections proportional to $t^1$ and $t^2$, the average energy losses
can be obtained in the same way, starting from the relevant expressions for  $\frac{d\Gamma}{d q_0}$.

\begin{figure}
    \centering
\includegraphics[width=0.49\textwidth]{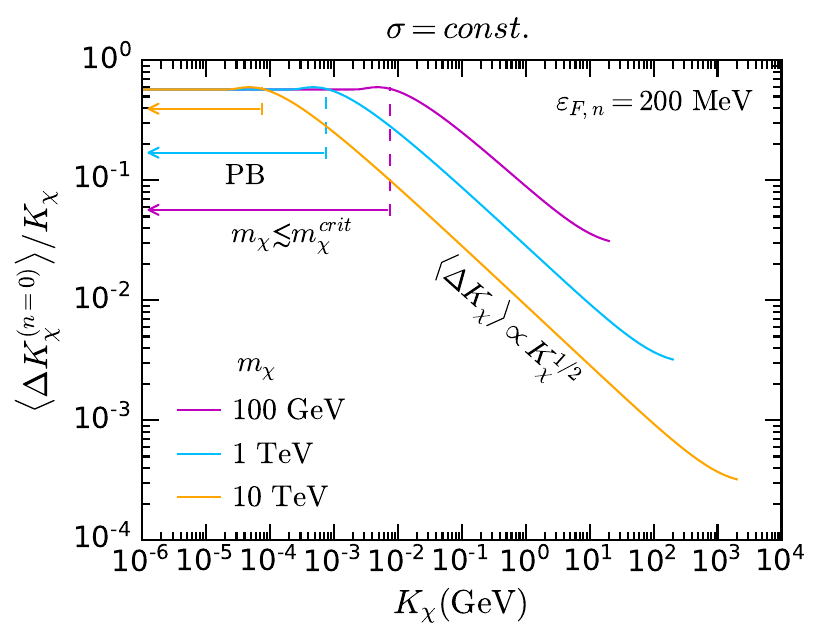}
\includegraphics[width=0.49\textwidth]{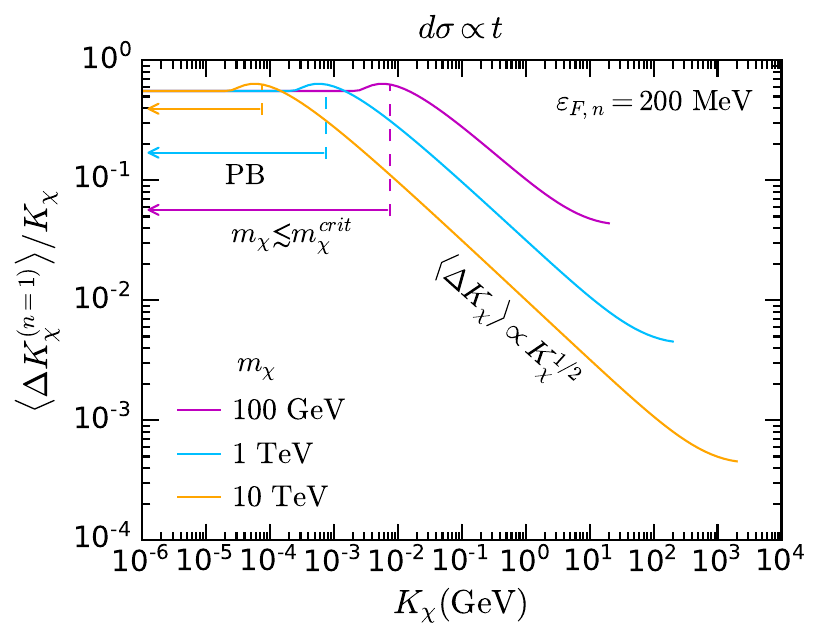}
    \caption{Average fraction of energy loss per DM-neutron collision for constant cross section (left) and $d\sigma\propto t$ (right) as a function of the DM kinetic energy. Horizontal arrows indicate the Pauli blocked (PB) regime, $m_\chi\lesssim m_\chi^{\rm crit}$. 
    } 
    \label{fig:Taven0}
\end{figure}

Figure~\ref{fig:Taven0} shows the average energy loss fraction per collision. We see the  $\langle \Delta K_\chi^{(n=0)} \rangle/K_\chi \propto K_\chi^{-1/2}$ scaling of Eq.~\ref{eq:aveElossn0largem}
in the $m_\chi \gtrsim m_\chi^{\rm crit}$ phase of the evolution, where the kinetic energy is driven down toward values where Pauli blocking eventually becomes active. The latter Pauli blocked phase is indicated by the horizontal arrows in Fig.~\ref{fig:Taven0}.

Moving to the case where Pauli blocking suppresses the scattering rate, $m_\chi\lesssim m_\chi^{\rm crit}$, the average energy loss per collision for the case of a constant cross section is
\begin{equation}
\langle \Delta K_\chi^{(n=0)} \rangle \sim 
\frac{4}{7}K_\chi. 
\label{eq:aveElossn0}
\end{equation}
The average energy loss now scales linearly with $K_\chi$ (the flat regions in Fig.~\ref{fig:Taven0}) in contrast to the $K_\chi^{1/2}$ dependence of Eq.~\ref{eq:aveElossn0largem}.
As the DM kinetic energy decreases, the average fraction of energy transferred to the targets progressively increases until $K_\chi$ no longer satisfies Eq.~\ref{eq:mucrit} and consequently the interaction rate becomes Pauli blocked.
As expected from Eq.~\ref{eq:mucrit}, the Pauli-suppressed region starts at higher kinetic energy for lower DM masses.

For interactions with $t$-dependent matrix elements, the average energy loss per collision also scales linearly with $K_\chi$ in the Pauli blocked regime. For $\Msq \propto t^n$, with $n=1,2$, we find 
\begin{equation}
\langle \Delta K_\chi^{(n=1)}\rangle \sim \frac{5}{9}K_\chi, \qquad 
\langle \Delta K_\chi^{(n=2)} \rangle \sim \frac{28}{55}K_\chi ,
\label{eq:aveElossn12}
\end{equation}
respectively, where we have used Eqs.~\ref{eq:intraten1} and \ref{eq:intraten2} for the interaction rates.
Note that the average energy loss fraction per collision exhibits similar behavior for all the interaction types considered,  
as seen by comparing the left and right panels of Fig.~\ref{fig:Taven0}. 
This is true in both the Pauli-blocked and non-blocked regimes.

\subsection{Thermalization timescale}
\label{sec:thermstandard}

Once a DM particle is captured, it becomes gravitationally bound to the NS and follows an orbit that may or may not lie completely within the NS. If the orbit lies partly outside the NS, subsequent scatterings will be required in order for the DM particle to lose enough energy so that the complete orbit lies within the NS. This is the first stage in the thermalization process. When estimating the amount of time needed for the DM orbit to lie completely within the star,  we find that this time is always much shorter than the full time required for DM to reach thermal equilibrium with the neutron targets. Consequently, this first step in the thermalization process can be safely neglected. This finding is in agreement with ref.~\cite{Garani:2018kkd}. 

We shall also assume that up-scattering of the DM to larger kinetic energy does not play an important role\footnote{Up-scattering refers to collisions with negative energy transfer $q_0<0$, such that the DM particle gains energy instead of losing it. When complete thermalization has been achieved, the rates for up-scattering and down-scattering must become equal, and hence we expect the up-scattering rates to become more significant as thermalization is approached.
If this were to be significant, our calculation below would underestimate the full thermalization time. As we shall see, this does not impact our final conclusions.}. These effects will become relevant as the DM approaches thermal equilibrium, increasing the thermalization time. We estimate that up-scattering will, at most, increase the thermal equilibrium time by $\mathcal{O}(10\%)$, and thus we neglect this correction.

For DM of mass much larger than the target mass, $m_\chi\gg \mbeff$, there is an additional stage in the thermalization process where either Pauli blocking plays no role, or the interaction rate has a different power law relationship with the temperature than those identified in Section~\ref{sec:energyloss}. These initial scatterings make a negligible contribution to the thermalization time, as  $\Gamma^{-}$ is a sharply decreasing function of the DM kinetic energy $K_\chi$.

Let us denote the number of initial collisions prior to reaching the Pauli blocked regime as $N_1$, and the number of additional collisions required for complete thermalization as $N_2$. For light DM, $m_\chi\lesssim \mbeff$, Pauli blocking affects the entire thermalization process, i.e. $N_1=0$. Let $K_N$ be the kinetic energy after $N$ scatterings. After $N_1+N_2$ collisions, the DM will reach the equilibrium temperature $T_{\rm eq}$, which can be written as
\begin{equation}
K_{N_1+N_2} = 
K_{N_1}\left(1-\frac{\langle \Delta K_\chi \rangle}{K_\chi}\right)^{N_2} = \Tstareq,  
\label{eq:Teq}
\end{equation}
where we have used the fact that the average fractional energy loss is the same in each collision. 
The thermalization time can then be defined as the sum of the average time between collisions, up until the final energy transfer is equal to $T_{\rm eq}$~\cite{Bertoni:2013bsa}
\begin{equation}
\tth = \sum_{n=0}^{N_2} \frac{1}{\Gamma^{-}(K_n)} \sim \sum_{n=N_1}^{N_2} \frac{1}{\Gamma^{-}(K_n)}.  
\label{eq:thermtime}
\end{equation}

For  $m_\chi \lesssim m_\chi^{\rm crit}$, the fraction of energy lost in the last few scatters is still a considerable fraction of the DM kinetic energy before the collision. Furthermore, these scatterings may take a considerably long time to occur, indicating that the process is discrete. 
As an example, consider thermalization to a temperature of $10^3$~K, for a DM particle of mass $m_\chi=1\TeV$ and constant cross section $\sigma_{n\chi} \sim 10^{-45}\cm^2$.\footnote{ For a wide DM mass range (GeV--TeV), this value is comparable to the cross-section that results in maximal capture, and hence we will use this cross-section as a reference value in some of the estimates that follow.} Hundreds of collisions are required to fully thermalize; the last 10 or so are spaced longer than a second apart; and the last couple are longer than $10 \kyr$ apart.

To compute the thermalization time, we numerically integrate Eq.~\ref{eq:diffGammaFiniteTemp} to obtain the interaction rate and the average energy lost in each collision for each of the EFT operators in Table~\ref{tab:operatorshe}. We find that the thermalization time for a particular interaction type scales according to the dominant power of the Mandelstam variable $t$ in the corresponding matrix element; see the last column of  Table~\ref{tab:operatorshe}. Thus, to understand how the thermalization times scale, it is enough to consider differential cross sections that are proportional to a given power of $t$, i.e. $d\sigma\propto t^n$, with $n=0,1,2$. 
Below, we present results for operators that depend only on $t$ (D1-4 in Table~\ref{tab:operatorshe}), and not on the centre of mass energy, $s$. In Appendix~\ref{sec:sdeptherm}, we outline the procedure used to obtain analytic expressions for those operators with an explicit dependence on $s$ (operators D5-10).

\begin{figure}
    \centering 
    \includegraphics[width=\textwidth]{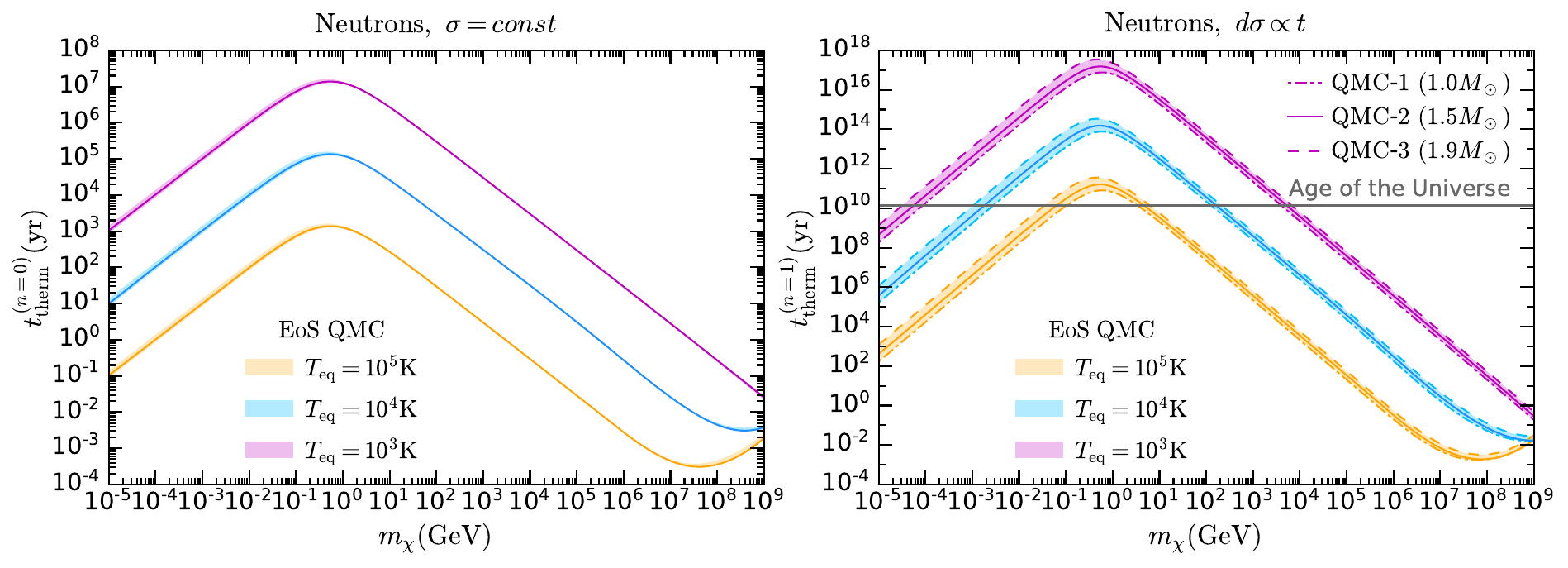}
    \includegraphics[width=.525\textwidth]{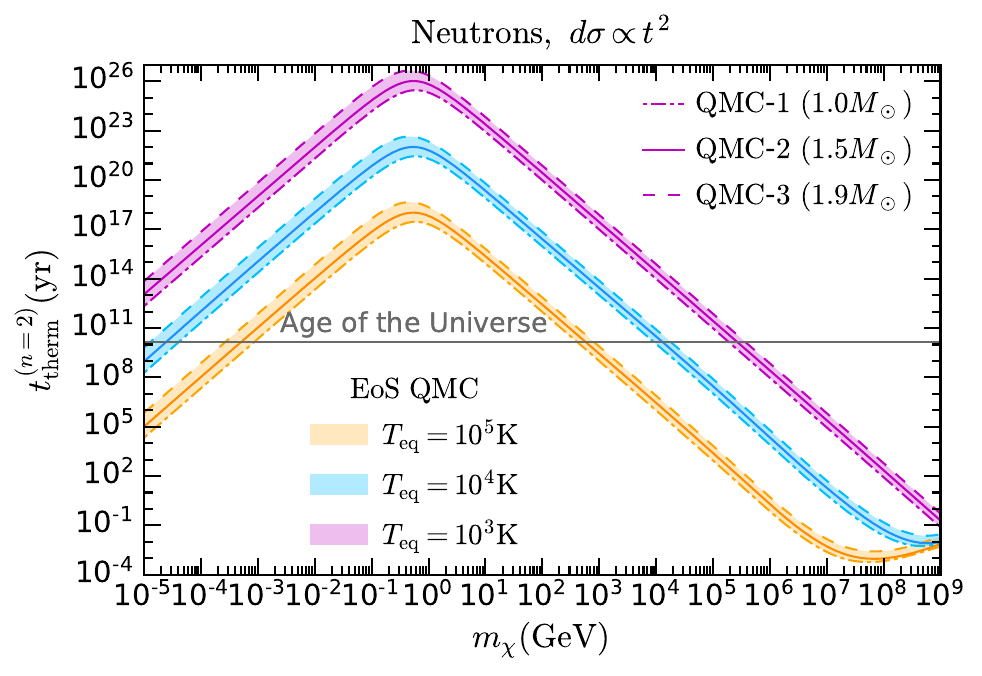}
    \caption{Thermalization time as a function of the DM mass for constant cross section (top left), $d\sigma \propto t$ (top right)  and $d\sigma\propto t^2$ (bottom). We have used the NS benchmark models in Table~\ref{tab:eos} and a reference cross section of $\sigma_{\rm n\chi}= 10^{-45}\cm^2$ close to the NS surface. Shaded regions indicate the variation with the choice of EoS: QMC-1 (dot-dashed), QMC-2 (solid) and QMC-3 (dashed). 
    \label{fig:thermtime}
    }
\end{figure}

In Figure~\ref{fig:thermtime}, we show the full numerical results for the thermalization time as a function of the DM mass for different equilibrium temperatures. It is clear that the power law scaling of the thermalization time with DM mass depends on whether $m_\chi$ is larger or smaller than the nucleon mass. In order to understand these results, we make use of the analytic approximations for the average DM energy loss derived in Section~\ref{sec:energyloss} and valid in the zero temperature limit. 

We begin by studying the Pauli blocked regime, $m_\chi \lesssim m_\chi^{\rm crit}$. 
In this case, the majority of the thermalization time is dictated by the final few scatters, for which the form factors are close to their value at zero momentum transfer.  These last collisions occur close to the NS centre, so we can take the target mass as constant and equal to the value at the centre of the star, $\mbeff(0)$. For the case of a constant DM-neutron cross section, $d\sigma \propto t^0$, the thermalization time can thus be obtained by using  Eqs.~\ref{eq:intraten0}, \ref{eq:Teq} and \ref{eq:aveElossn0} in Eq.~\ref{eq:thermtime}. This leads to
\begin{equation}
\tthn{0} \sim   \frac{147}{16 }\frac{\pi^2 m_\chi}{ \left(\mbeff(0) + m_\chi\right)^2}\frac{1}{\sigma_{i\chi}^{n=0}}\frac{1}{T_{\rm eq}^2}, 
\label{eq:thermtimen0}
\end{equation} 
where $\sigma_{i\chi}^{n=0}$ is the DM-baryon cross section. 
The scaling of this expression with $\Tstareq$, DM mass, and DM-target cross sections 
agrees with ref.~\citep{Bertoni:2013bsa}. 
Numerically we obtain
\begin{equation}
    \tthn{0} \sim \begin{dcases}
        4.4\times 10^{6}\yrs \, \left(\frac{m_\chi}{10\MeV}\right)\left(\frac{10^{-45}\cm^2}{\sigma_{i\chi}^{n=0}}\right)\left(\frac{10^3K}{T_{\rm eq}}\right)^{2},\quad m_\chi \ll \mbeff(0)\\
        9.7\times 10^{6}\yrs \, \left(\frac{10\GeV}{m_\chi}\right)\left(\frac{10^{-45}\cm^2}{\sigma_{i\chi}^{n=0}}\right)\left(\frac{10^3K}{T_{\rm eq}}\right)^{2},\quad m_\chi \gg \mbeff(0)
    \end{dcases} \label{eq:t_therm_0}
\end{equation}
where we have set $\mbeff(0) = 0.5\,m_n$. 
The $m_\chi$ dependence of these expressions explains the features observed in the top left panel of Fig.~\ref{fig:thermtime}. For $m_\chi \ll \mneff(0)$, the thermalization time scales with the DM mass as the number of scatterings needed for thermalization increases. Conversely, for $m_\chi \gg \mneff(0)$, $\tthn{0}$ is inversely proportional to $m_\chi$ due to the reduced Pauli blocking, explaining the change of slope around the value of the neutron effective mass in the NS centre.

We repeat the same analysis for cross section proportional to higher powers of $t$, i.e., $d\sigma \propto t^n$ with $n=1,2$. Using Eqs.~\ref{eq:intraten1}, \ref{eq:intraten2} and \ref{eq:aveElossn12}, we find
\begin{align}
    \tthn{1} &\sim 14 
    \frac{\pi^2 m_\chi^2 (\mbeff(0))^2}{\left(\mbeff(0) + m_\chi\right)^2 \left((\mbeff(0))^2+ m_\chi^2\right)}\frac{1}{\sigma_{i\chi}^{n=1}}\frac{1}{T_{\rm eq}^3} \frac{1-B(\Rstar)}{B(\Rstar)},  
    \label{eq:thermtimen1}\\
    \tthn{2} &\sim 17 
    \frac{\pi^2 m_\chi^3 (\mbeff(0))^4}{\left(\mbeff(0) + m_\chi\right)^2 \left((\mbeff(0))^2+ m_\chi^2\right)^2}\frac{1}{\sigma_{i\chi}^{n=2}}\frac{1}{T_{\rm eq}^4} \left[\frac{1-B(\Rstar)}{B(\Rstar)}\right]^2, 
    \label{eq:thermtimen2}
\end{align}
where the factors involving $B(R_\star)$ arise from fixing the cross section to its value at the surface; 
see Appendix~\ref{sec:pauliblockingle} for details.

To gain insight into the order of magnitude of these thermalization times, we set  $B(\Rstar)=0.5$ and $\mbeff(0) = 0.5\,m_n$, yielding
\begin{align}
    \tthn{1} &\sim \begin{dcases}
        2.5\times 10^{14}\yrs \, \left(\frac{m_\chi}{10\MeV}\right)^2\left(\frac{10^{-45}\cm^2}{\sigma_{i\chi}^{n=1}}\right)\left(\frac{10^3K}{T_{\rm eq}}\right)^{3},\quad m_\chi \ll \mbeff(0)\\
        3.9\times 10^{15}\yrs \, \left(\frac{10\GeV}{m_\chi}\right)^2\left(\frac{10^{-45}\cm^2}{\sigma_{i\chi}^{n=1}}\right)\left(\frac{10^3K}{T_{\rm eq}}\right)^{3},\quad m_\chi \gg \mbeff(0)
    \end{dcases}
    \label{eq:t_therm_1} \\
    \tthn{2} &\sim \begin{dcases}
        1.1\times 10^{23}\yrs \, \left(\frac{m_\chi}{10\MeV}\right)^3\left(\frac{10^{-45}\cm^2}{\sigma_{i\chi}^{n=2}}\right)\left(\frac{10^3K}{T_{\rm eq}}\right)^{4},\quad m_\chi \ll \mbeff(0)\\
        1.2\times 10^{24}\yrs \, \left(\frac{10\GeV}{m_\chi}\right)^3\left(\frac{10^{-45}\cm^2}{\sigma_{i\chi}^{n=2}}\right)\left(\frac{10^3K}{T_{\rm eq}}\right)^{4},\quad m_\chi \gg \mbeff(0).
    \end{dcases}
    \label{eq:t_therm_2}
\end{align} 
As anticipated, we see that the momentum-suppressed cross sections translate into significantly longer thermalization times than for the case of a constant (unsuppressed) cross section.
These expressions also allow us to understand the dependence of $\tth$ on the DM mass.
For $d\sigma\propto t^n$, the thermalization time scales as $m_\chi^{n+1}$ for $m_\chi \ll \mneff(0)$, and as the inverse of this quantity for $m_\chi \gg \mneff(0)$.

The choice of EoS has a small but non-negligible impact on the thermalization time, as indicated by the widths of the shaded regions in Fig.~\ref{fig:thermtime}. 
For a constant cross section, we observe almost no variation in $\tth$ with the NS configuration, except for the $m_\chi \lesssim m_n$ region. This is due to the dependence of $\mneff(0)$ on the NS model; see Table~\ref{tab:eos} and Eq.~\ref{eq:thermtimen0}.
For cross sections $d\sigma\propto t^n$, with $n=1,2$, the dependence on $B(\Rstar)$ in Eqs.~\ref{eq:thermtimen1} and \ref{eq:thermtimen2} adds an extra dependence on the choice of NS model. For these momentum-suppressed interactions, DM requires more time to reach an equilibrium temperature in heavier NSs. This is due to the combination of two effects: the effective mass of the targets in the centre of the NS is smaller in more massive NS configurations, while $B(\Rstar)$ increases. Nonetheless, the dependence on NS configuration remains relatively mild. 


We now turn to the $m_\chi \gtrsim \mcrit$ regime, which is observed only for temperatures above $10^4\K$ in Fig.~\ref{fig:thermtime}. This regime change is indicated by the change of slope that occurs at large DM masses, clearly evident for  $T_{eq}=10^5\K$ (orange)  at a DM of mass $m_\chi\gtrsim5\times10^7\GeV$.  
In this regime, the energy lost in each collision is a tiny fraction of the initial DM kinetic energy, and the time between scatterings is of order a fraction of a second. This indicates that a continuous approximation becomes more appropriate than a discrete sum to estimate $\tth$.
In this case, the momentum-dependent part of the form factor, $F(t)$, will be relevant only at the beginning of the thermalization process and become less and less relevant as the average momentum transfer decreases in each subsequent scatter.\footnote{We have numerically verified that the $t$-dependent form factors do not alter the results in any significant manner.} It is these low momentum-transfer collisions that dominate the thermalization time.
For a constant cross section ($n=0$), in the zero temperature approximation, we obtain 
(see Appendix~\ref{sec:thermsuperheavy} for details)
\begin{equation}
    \tthn{0} \sim \frac{9 \pi^2 m_\chi}{8 (\mbeff(0))^2 \kinFi^2 \sigma_{i\chi}^{n=0} }\log\left[\frac{m_\chi}{T_{\rm eq}}\left(\frac{1}{\sqrt{B(\Rstar)}}-1\right)\right]. 
\label{eq:tthemheavy0text}
\end{equation}
In this super heavy DM mass regime, we see that the thermalization time is an increasing function of $m_\chi$.

It is worth remarking that for a constant DM-neutron cross section (top left panel) thermalization will always occur within the age of the Universe. However, this is not true for momentum-suppressed cross sections, for a range of DM masses. Specifically, for $T_{\rm eq}=10^3\K$ and the assumed reference cross section,  DM of mass $m_\chi\lesssim 10 \TeV$ ($m_\chi\lesssim 1 \PeV$)  will not have enough time to  thermalize  for $d\sigma \propto t$ ($d\sigma \propto t^2$).
Importantly, however, we shall see below that even when full thermalization takes longer than the age of the Universe, the majority of the kinetic energy is deposited on a much shorter timescale.

Finally, we must incorporate the fact that DM will scatter with various baryonic species in the NS rather than just the neutrons. To do this, we combine the thermalization times for scattering from each individual species in an appropriate way. Specifically, we sum the inverse single-species thermalization times, weighted by their relative abundance at the centre of the NS, such that
\begin{equation}
    \frac{1}{t_{\text{therm},\; \text{tot}}} = \sum_i \frac{Y_i(0)}{t_{\text{therm},\;i}},
    \label{eq:tthermtot}
\end{equation}
where $Y_i(0)$ is the abundance of the species in the centre of the NS, and the sum runs over all possible baryons. For the case of the heaviest NS we consider, 1.9 $\Msun$, this includes the $\Lambda^0$, $\Xi^0$ and $\Xi^-$ hyperons. The resulting thermalization time then lies between the fastest and slowest single-species times, as is expected.

\section{Capture-Annihilation Equilibrium}
\label{sec:CAEquilibrium}

The captured DM will accumulate in the centre of the NS, where it will begin to annihilate. The annihilation rate will grow until sufficient time has elapsed for the capture and annihilation processes to reach equilibrium. In this limit, the total amount of DM in the NS is maximized, and will remain constant. Once this occurs, annihilation efficiently deposits the DM mass-energy into the star.

Let us begin by assuming that the dark matter has fully thermalized. After reviewing the standard capture-annihilation equilibrium calculation, we will relax this assumption to consider the more general case of partially thermalized dark matter and derive new expressions that hold in that scenario. Importantly, we shall see that capture-annihilation equilibrium, and hence efficient annihilation can occur without full thermalization.

\subsection{Capture-annihilation equilibrium of thermalized dark matter}

The thermalized DM will collect within an isothermal sphere at the centre of the NS where it will begin to annihilate. The efficiency of the annihilation will depend on the volume of this sphere, which is expected to be very small for the DM masses we consider. 
Very close to the centre of the NS, the density does not vary significantly and can be taken to be constant. Then, working in the weak field approximation such that $B(r) \sim  1+2\Phi(r)$, 
we can obtain the gravitational potential inside the NS,
\begin{align}
\Phi(r) & = -\int_r^\infty \frac{G M_\star(r')}{r'^2}dr' 
\approx \frac{2}{3}\pi G \rho_c r^2, 
\end{align}
where $\rho_c$ is the central density of the NS. 
The  number density of DM particles that have thermalized to a temperature $\Tstareq$ as a function of radius will then be given by a Maxwell-Boltzmann distribution
\begin{align}
n_{\chi}(r) & \simeq n_0 \exp\left[ -\frac{m_\chi\Phi(r)}{T_{\rm eq}}\right] 
= \frac{N_\chi}{\pi^{3/2}  r_{\rm iso}^3}\exp\left( -\frac{r^2}{ r_{\rm iso}^2}\right), 
\end{align}
where $N_\chi$ is the total number of DM particles within the isothermal sphere, and $r_{\rm iso}$ is the radius of the DM isothermal sphere. 
Applying the viral theorem leads to the following expression for $r_{\rm iso}$, 
\begin{align}
r_{\rm iso} & = \sqrt{\frac{3 \Tstareq}{2\pi G m_\chi \rho_c}} \nonumber\\
& \approx  0.26\,\text{m}\left[\left(\frac{T_{\rm eq}}{10^3\,\text{K}}\right)\left(\frac{1\GeV}{m_\chi}\right)\left(\frac{8\times10^{14}\,\textrm{g}\,\text{cm}^{-3}}{\rho_c}\right)\right]^{1/2}.
\end{align}
The total number of DM particles enclosed in this sphere is then
\begin{equation}
    N_\chi    \simeq  4\pi\int dr \, r^2n_{\chi}(r),
\end{equation} 
and the velocity distribution of the thermalized DM is  given by
\begin{equation}
f_{\rm MB}(v_\chi)  = 4\pi \left(\frac{m_\chi}{4 \pi \Tstareq}\right)^{3/2} v_\chi^2 \exp\left[ -\frac{m_\chi v_\chi^2}{4 \Tstareq} \right].
\end{equation}

In the absence of evaporation, which can safely be neglected for $m_\chi \gtrsim 1.17\times 10^{-8}\GeV$ for a Gyr old NS with core temperature $\sim 10^3\K$~\cite{Bell:2020lmm}, the time evolution of the total number of DM particles present inside the NS is governed by
\begin{equation}
    \frac{dN_\chi}{dt} = C - A N_\chi^2 \label{eq:ndm}.
\end{equation}
Here $C$ is the capture rate and  $A$ is related to the DM annihilation rate, $\Gamma_{\rm ann}$, through
\begin{equation}
    \Gamma_{\rm ann} =  \frac{1}{2}A N_\chi^2,
\end{equation}
where
\begin{equation}
    A = \frac{\langle \sigma_\text{ann}v_\chi \rangle}{N_\chi^2} \int n_\chi^2(r) d^3 r \simeq \frac{\langle \sigma_\text{ann} v_\chi\rangle}{(2\pi)^{3/2} r_{\rm iso}^3}, \label{eq:annrate}
\end{equation}
and $\langle \sigma_{\rm ann} v_\chi \rangle$ is the
thermally averaged DM annihilation cross section. These are given in Table~\ref{tab:annCS} for the EFT interactions we consider, where $\langle v^2_\chi \rangle = v_{\rm th}^2 =  6 \Tstareq/ m_\chi$.

We note that the cross sections shown in Table~\ref{tab:annCS} are quark-level expressions. For most of the mass range of interest, these provide excellent approximations to the hadron-level annihilation cross sections, provided we impose a lower bound on the DM mass for which an annihilation channel is open, taken to be the pion mass. See Appendix~\ref{sec:quarkhadron} for details.

\begin{table}[t]
    \centering
    \begin{tabular}{|c|c|l|}
    \hline
         Name & Operator & $\langle \sigma_{ann} v_\chi\rangle$\\
        \hline 
         D1  & $\bar\chi  \chi\;\bar q  q $ & 
         $\frac{3 m_\chi^2}{8\pi\Lambda^4}\sum_q y_q^2 \left( 1 - \frac{m_q^2}{m_\chi^2}\right)^{3/2} v_{\rm th}^2 $\\ \hline
         D2  &  $\bar\chi \gamma^5 \chi\;\bar q q $ &  \large $\frac{3 m_\chi^2}{2\pi \Lambda^4}\sum_q y_q^2 \sqrt{1 - \frac{m_q^2}{m_\chi^2}} \left[ \left( 1 - \frac{m_q^2}{m_\chi^2}\right) + \frac{3}{8}\frac{m_q^2}{m_\chi^2}v_{\rm th}^2\right]$\\ \hline
         D3  & $\bar\chi \chi\;\bar q \gamma^5  q $  & $\frac{3 m_\chi^2}{8\pi\Lambda^4}\sum_q y_q^2 \sqrt{ 1 - \frac{m_q^2}{m_\chi^2}} v_{\rm th}^2$\\ \hline
         D4  & $\bar\chi \gamma^5 \chi\; \bar q \gamma^5 q $ & $ \frac{3 m_\chi^2}{2\pi\Lambda^4}\sum_q y_q^2 \sqrt{1 - \frac{m_q^2}{m_\chi^2}}\left[ 1 + \frac{m_q^2}{8(m_\chi^2 - m_q^2)} v_{\rm th}^2\right]$\\ \hline
         D5  & $\bar \chi \gamma_\mu \chi\; \bar q \gamma^\mu q$ & $\frac{3 m_\chi^2}{2\pi \Lambda^4}\sum_q \sqrt{1 - \frac{m_q^2}{m_\chi^2}} \left[ \left( 2 +\frac{m_q^2}{m_\chi^2}\right) + \left( \frac{-4m_\chi^4 +2m_q^2 m_\chi^2 +11m_q^4}{24m_\chi^2 ( m_\chi^2 - m_q^2)} \right) v_{\rm th}^2 \right]$\\ \hline
         D6  & $\bar\chi \gamma_\mu \gamma^5 \chi\; \bar  q \gamma^\mu q $ & $\frac{m_\chi^2}{4\pi \Lambda^4}\sum_q \sqrt{1 - \frac{m_q^2}{m_\chi^2}} \left[ 2 +\frac{m_q^2}{m_\chi^2}\right] v_{\rm th}^2$\\ \hline
         D7  & $\bar \chi \gamma_\mu  \chi\; \bar q \gamma^\mu\gamma^5  q$  & $ \frac{3 m_\chi^2 }{\pi\Lambda^4}\sum_q \sqrt{1 - \frac{m_q^2}{m_\chi^2}} \left[ \left( 1 - \frac{m_q^2}{m_\chi^2} \right) - \frac{1}{24}\left( 2 - 11\frac{m_q^2}{m_\chi^2}\right) v_{\rm th}^2 \right]$ \\ \hline
         D8  & $\bar \chi \gamma_\mu \gamma^5 \chi\; \bar q \gamma^\mu \gamma^5 q $ & $ \frac{3 m_\chi^2}{2\pi \Lambda^4} \sum_q \sqrt{1 - \frac{m_q^2}{m_\chi^2}} \left[ \frac{m_q^2}{m_\chi^2} + \left( \frac{8m_\chi^4 - 28m_\chi^2 m_q^2 + 23 m_q^4}{24 m_\chi^2(m_\chi^2 - m_q^2)} \right) v_{\rm th}^2 \right] $ \\ \hline
         D9  & $\bar \chi \sigma_{\mu\nu} \chi\; \bar q \sigma^{\mu\nu} q $  & $ \frac{6 m_\chi^2}{\pi\Lambda^4} \sum_q \sqrt{1 - \frac{m_q^2}{m_\chi^2}} \left[ \left( 1 + 
         2\frac{m_q^2}{m_\chi^2}\right) -\left( \frac{2m_\chi^4 +17m_q^2 m_\chi^2 -28m_q^4}{24m_\chi^2(m_\chi^2 - m_q^2) }\right) v_{\rm th}^2\right] $ \\ \hline         
         D10 & $\bar \chi \sigma_{\mu\nu} \gamma^5\chi\; \bar q \sigma^{\mu\nu} q$ & $\frac{6 m_\chi^2}{\pi\Lambda^4} \sum_q \sqrt{1 - \frac{m_q^2}{m_\chi^2}} \left[ \left( 1 - \frac{m_q^2}{m_\chi^2}\right)  -\frac{1}{24}\left(2 -17\frac{m_q^2}{m_\chi^2}\right) v_{\rm th}^2 \right] $\\ \hline
    \end{tabular}
    \caption{Thermally averaged annihilation cross sections $\sigmav$ for the dimension 6 EFT operators, expanded to second order in $v_\chi$. The $y_q$ factors are the fermion Yukawa couplings~\cite{Zheng:2010js}.}
    \label{tab:annCS} 
\end{table}

The solution to Eq.~\ref{eq:ndm} in terms of the capture and annihilation rates is
\begin{equation}
    N_\chi(t) = \sqrt{\frac{C}{A}}\tanh\left(\sqrt{CA} \,t\right).\label{eq:Noft}
\end{equation}
Ultimately, we are interested in the behavior of Eq.~\ref{eq:Noft} at late stages in the NS evolution, i.e., for $t\rightarrow \tstar$, where $\tstar$ is the age of the NS, which we take to be $\sim 1\Gyr$. In this limit, the hydrostatic NS structure, and hence $C$, are not expected to change with time. Of particular interest is whether or not an equilibrium is reached between the capture and annihilation rates. Such a state is reached for timescales greater than 
\begin{equation}
     t_{\rm eq} = \frac{1}{\sqrt{C A}}.
     \label{eq:teqold}
\end{equation}
For earlier times, $ t < t_{\rm eq}$, one can neglect the loss of DM particles due to annihilation, leaving $N_\chi\sim C t$.

\subsection{Capture-annihilation equilibrium of partially-thermalized dark matter}

The standard calculation of the annihilation rate, using Eq.~\ref{eq:annrate}, assumes that the DM has thermalized, i.e., $t > t_{\rm therm}$. 
If thermalization has not been achieved by a time $t\sim\tstar<\tth$, the DM kinetic energy distribution will peak around the lowest temperature that DM has had enough time to reach. This is given by 
\begin{equation}
K_\chi \sim T_{\rm eq} \left(\frac{\tth+\tstar}{\tstar}\right)^{\frac{1}{2+n}},
\label{eq:lowtemp}
\end{equation}
where $n$ is the exponent of the dominant $t^n$ term in the differential cross section, $d\sigma \propto t^n$, as given in the last column of Table~\ref{tab:operatorshe}. See Appendix~\ref{sec:minTempDerivation} for details. 
We can then find the radius of the DM distribution (which is no longer isothermal) and the $\sigmav$ corresponding to the peak of the energy distribution $K_\chi$. 
We obtain $A$ via the replacement  
\begin{equation}
    A \rightarrow A \left(\frac{T_\text{eq}}{K_\chi}\right)^{\alpha} = 
   A \left(\frac{\tstar}{t_\text{therm}+\tstar}\right)^{\frac{\alpha}{2+n}},  
\end{equation}
where $\alpha=3/2$ for $s$-wave annihilation, and $\alpha=1/2$ for $p$-wave. 
Making this replacement in Eq.~\ref{eq:teqold} leads to a capture-annihilation equilibrium time of
\begin{equation}
 t_{\rm eq} = \frac{1}{\sqrt{C A}} \left(\frac{t_\text{therm}+\tstar}{\tstar}\right)^{\frac{\alpha}{2(2+n)}}. \label{eq:teq} 
\end{equation}

No previous estimate of the capture-annihilation equilibrium time has considered the case of partially thermalized DM. If thermalization has not been achieved, the additional factor in Eq.~\ref{eq:teq}, compared to Eq.~\ref{eq:teqold}, increases the equilibrium time. (Or, equivalently, increases the cross sections required to reach equilibrium within a specified time.) However, it is critical to realize that $t_\text{eq}$ can be shorter than $t_\text{therm}$. 
In fact, it is possible for annihilation to occur efficiently, even if complete thermalization never occurs. In this scenario, we must use Eq.~\ref{eq:teq}.

Assuming the DM is captured at the geometric limit, we arrive at the following result for our benchmark NS QMC-2
\begin{eqnarray}
    t_{\rm eq}  &\sim&  4\times10^{-6}\yr\left(\frac{100\GeV}{m_\chi}\right)^{\frac{1}{4}}\left(\frac{10^{-26}\cm^3 \s^{-1}}{\langle\sigma_\text{ann} v_\chi\rangle}\right)^{\frac{1}{2}} \left(\frac{1\GeV\cm^{-3}}{\rho_\chi}\right)^{\frac{1}{2}}\left(\frac{T_{\rm eq}}{10^3\K}\right)^{\frac{3}{4}} 
    \nonumber \\ &&\times
    \left(\frac{t_{\rm therm}+\tstar}{\tstar}\right)^{\frac{\alpha}{2(2+n)}}.
\end{eqnarray}
Comparing this expression with the thermalization times in the previous section, we anticipate that  $t_\text{eq}$ will typically be shorter than $t_\text{therm}$, often by many orders of magnitude.

\section{Neutron Star Heating Timescales}
\label{sec:results}

\subsection{Kinetic heating timescale}
\label{subsec:KinHeating}

It has commonly been assumed that the DM kinetic energy deposition occurs instantaneously. However, it is not immediately obvious that this is true. In particular, for scattering interactions suppressed by powers of the momentum transfer, $t$, full thermalization can take longer than the age of the Universe. We must therefore determine whether a {\it significant fraction} of the initial kinetic energy can be deposited on a shorter timescale.

To quantify the timescale on which kinetic heating takes place, we define $t_{\rm kin}$ to be the time required for a DM particle to lose 99\% of its maximum kinetic energy, $K_\chi = m_\chi (1/\sqrt{B(0)} - 1)$.  
This time is calculated in the same way as the thermalization time, while also keeping track of the time the DM spends outside the star along its orbit, i.e., the initial stage of the thermalization process that was neglected in Section~\ref{sec:thermstandard}. For simplicity, the DM particle orbits are taken to be linear, passing through the centre of the star, with the radial extent calculated using the geodesic equations. We expect an ${\cal O}(1)$ correction to our results when considering circular orbits.
Additionally, we randomize the radial position in the NS where the DM interacts with a target. 

\begin{figure}[t] 
    \centering  \includegraphics[width=.5\textwidth]{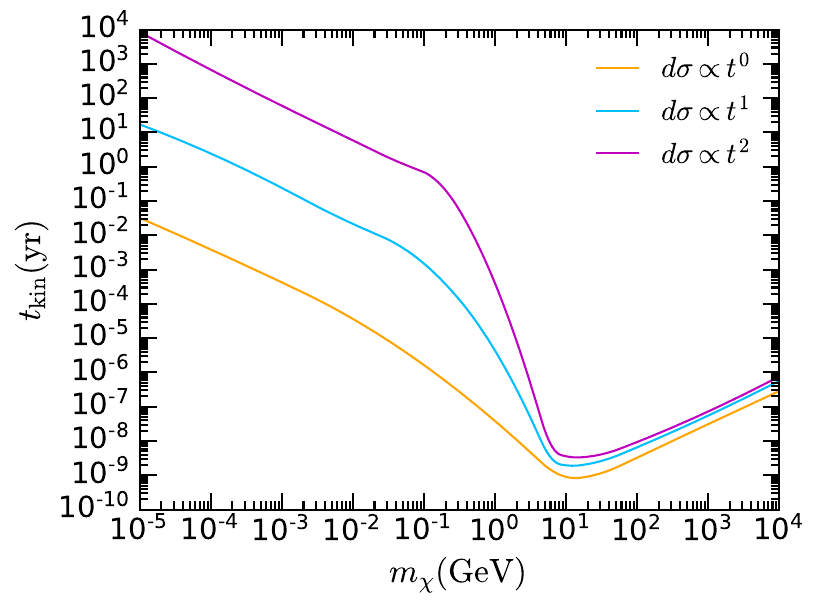}
    \caption{Timescale on which the DM deposits $99\%$ of its initial kinetic energy in the NS. We have assumed a NS with configuration QMC-2, and a DM-neutron scattering cross section of $\sigma_{n\chi}=10^{-45}\cm^2$ at the surface of the star. 
    }
    \label{fig:kinheattimes}
\end{figure}

Figure~\ref{fig:kinheattimes} shows the time required for kinetic heating to be achieved, assuming DM-neutron interactions of the form $d\sigma \propto t^n$, with cross sections normalized to $\sigma_{n\chi} = 10^{-45}\cm^2$ at the surface of the star. As the location of each interaction is randomized, these results are obtained by averaging over several simulations for each DM mass. For light DM, $t_{\rm kin}$ decreases with increasing DM mass, due to the decreased effects of Pauli blocking, with the change of slope at $m_\chi \sim 0.1\GeV$ indicating the point where Pauli blocking affects only a fraction of the total process.
For masses $\gtrsim 10\GeV$, Pauli blocking is not relevant this part of the thermalization process, and hence the $t_{\rm kin}$ monotonically increases with the DM mass, as was seen in the thermalization of super-heavy DM. 

Fig.~\ref{fig:kinheattimes} illustrates two key facts. First, $t_{\rm kin}$ differs by orders of magnitude for the different cross section types, $d\sigma\propto t^n$, with larger values of $t_{\rm kin}$ for the most highly momentum-suppressed interactions, as expected. Second, and importantly, the kinetic heating occurs relatively quickly for all interaction types, on timescales  much shorter than a typical NS age. Indeed, for the case of a constant cross section, $\tkin$ is much shorter than a year.

\subsection{Annihilation heating timescale}

\begin{figure}[t]
\centering    
\includegraphics[width=\textwidth]{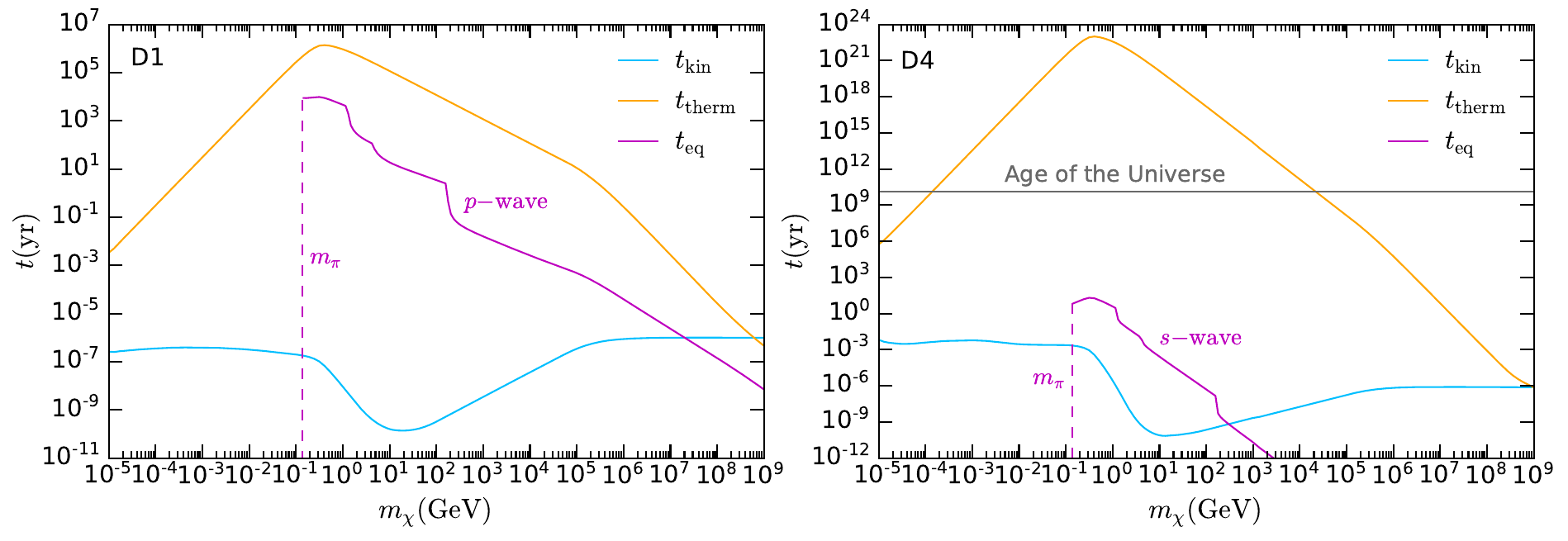}
    \caption{Timescales for kinetic heating (blue), thermalization (orange) and capture-annihilation equilibrium (magenta), for operators D1 (left) and D4 (right). The operator D1 has an unsuppressed scattering cross section and a $p$-wave suppressed annihilation cross section, while $D4$ has a $q_\text{tr}^4$ suppressed scattering cross section and an unsuppressed ($s$-wave) annihilation cross section.
    The interaction strength has been chosen to give maximal capture. (Specifically, we used $\Lambda_q$ values corresponding to a capture cross section at the geometric limit, assuming scattering with the neutron targets in the QMC-2 NS.)   
    }
    \label{fig:timescales}
\end{figure}

Figure~\ref{fig:timescales} shows all relevant timescales for DM-induced heating of old neutron stars. These timescales have been calculated considering DM scattering off the neutron targets in the benchmark NS QMC-2, for the case of maximal capture, $f = 1$ (i.e., we have set the EFT parameter $\Lambda_q$ as required to achieve capture at the geometric limit).
We show these results for two indicative operators: The scalar-scalar interaction D1 (left), which has a $p$-wave suppressed annihilation cross section, and the pseudoscalar-pseudoscalar operator D4 (right), which has an $s$-wave annihilation cross section. 

As anticipated, capture-annihilation equilibrium takes longer to achieve for the velocity-suppressed $p$-wave annihilation cross section than for the $s$-wave.  Nonetheless,  
equilibrium (and hence maximal annihilation heating) is reached relatively quickly in both cases, on timescales of $10^4$ years for the scalar interaction, and even quicker for the pseudoscalar, well within the age of a typical NS.

For both interaction types, the kinetic and annihilation heating contributions are both realized on timescales much shorter than that required for full thermalization. If the scattering cross section is momentum suppressed (as with the $d\sigma \propto t^2=q_\text{tr}^4$ dependence for D4), the thermalization time is increased; if the annihilation cross section is velocity suppressed (as with the $p$-wave annihilation cross section for D1) the capture-annihilation equilibrium time is increased.

Finally, note that there are parameters for which the annihilation timescale $\teq$ is shorter than the kinetic heating timescale $\tkin$. In this case, the annihilation process deposits the DM mass energy and any remaining kinetic energy, which is carried by the annihilation products. Therefore, the minimum time required for DM to deposit {\it all} of its energy, both kinetic and rest-mass, is $\teq$.

\subsection{Neutron star heating sensitivity for various interaction types}   
\label{ref:heatingresults}

\begin{figure}[t]
    \centering
    \includegraphics[width=0.622\textwidth]{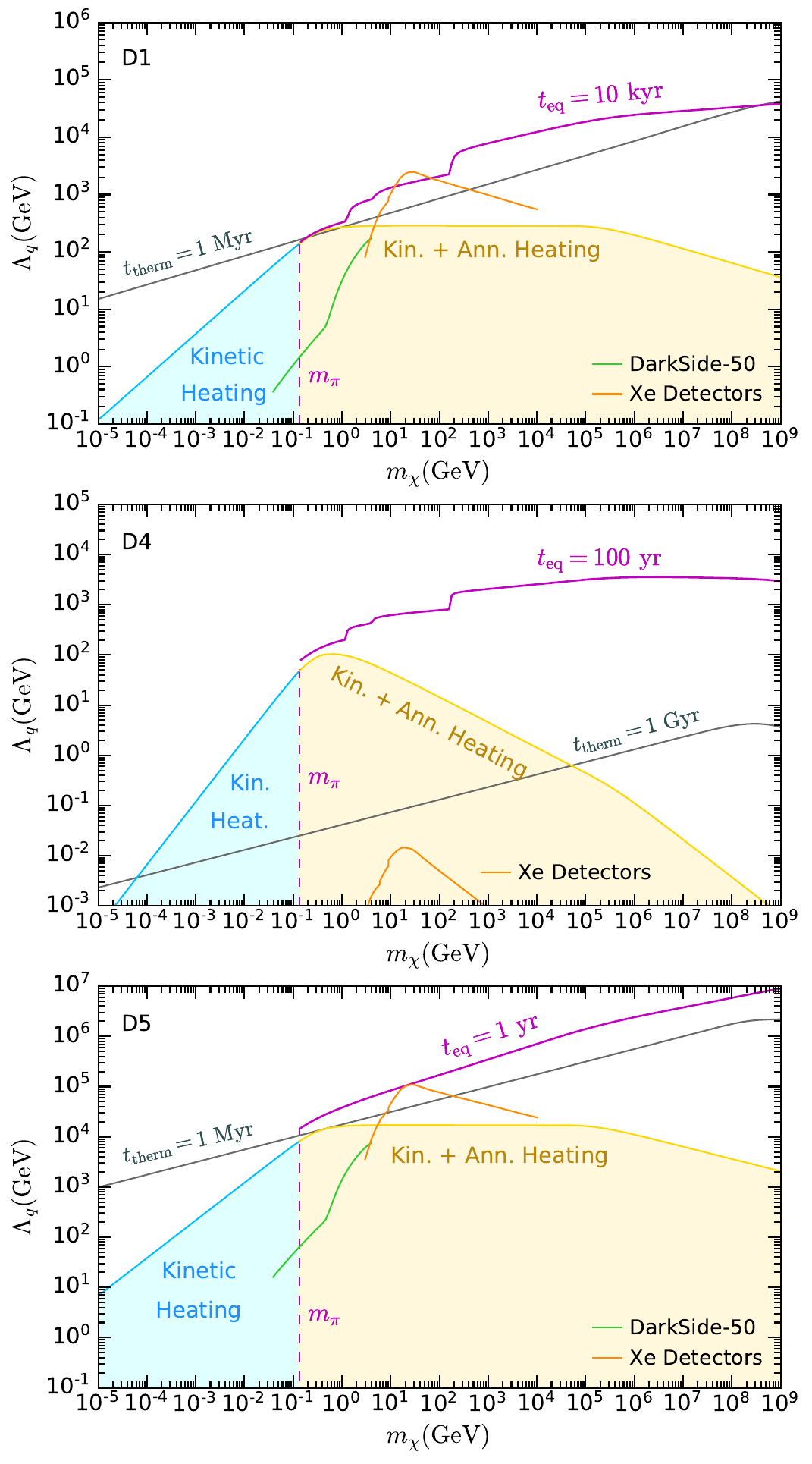}   
    \caption{
    Projected NS heating sensitivity for maximal capture efficiency, for DM-baryon interactions described by operators D1, D4 and D5.  We have used the QMC-2 ($1.5\Msun$) benchmark NS configuration.
    We show the regions where DM kinetic and annihilation heating both contribute to the NS luminosity (yellow) and where kinetic heating alone contributes (light blue).
    Contour lines for capture-annihilation equilibrium (magenta) and thermalization (grey) are shown for indicative timescales.    
     Lower limits on $\Lambda_q$ from leading  direct detection experiments~\cite{DarkSide:2022dhx,Aprile:2020thb,PandaX-4T:2021bab,LUX-ZEPLIN:2022xrq} are also shown.  
    }
    \label{fig:NS_heating}
\end{figure}

We now examine the regions of parameter space where maximal heating can be achieved, for DM-hadron interactions described by the four-fermion operators of Table~\ref{tab:operatorshe}. As the extent of DM-induced heating depends on how efficiently the DM is captured, it is clear that maximal heating corresponds to the case of $f=1$ in Eqs.~\ref{eq:kinenergy} and~\ref{eq:massheating}, i.e., when the dark matter scattering cross section is at or above the geometric limit.

In Figure~\ref{fig:NS_heating}, we show the parameters for which the DM deposits its entire kinetic and rest mass energy (yellow region) or only its full kinetic energy (light blue region).  Above these shaded regions, the value of $\Lambda_q$ is too large (and hence the scattering cross section is too small) for maximal capture.
The overall shape of the shaded regions is dictated by the behaviour of the capture rate. For sub-GeV DM, Pauli blocking suppresses the capture rate, and so smaller $\Lambda_q$ values are required to reach the geometric limit. At large DM mass, $\gtrsim 10^5\GeV$, the capture rate is suppressed because a single collision does not transfer enough energy to result in capture.

We see from the $\teq$ contours (magenta) that capture-annihilation equilibrium, and hence full annihilation heating, is achieved on a timescale much shorter that the NS lifetime, which we take to be $t_\star\sim 1\Gyr$. 
We do not show the contours for the kinetic heating timescale, $t_\mathrm{kin}$, as this is significantly shorter than the age of the star.  Moreover, for masses where DM annihilation to hadrons is possible, $m_\chi>m_\pi$, it is not necessary for kinetic heating to occur before capture-annihilation equilibrium is established, as the total kinetic plus mass energy will be deposited when the DM annihilates. 
For completeness, we include the contours of the thermalization time (grey). We stress again that the DM does not need to fully thermalize to achieve maximal heating.

To highlight how the sensitivity varies for different interaction types, we show results for operators D1, D4 and D5 in Figure~\ref{fig:NS_heating}, with the remaining operators presented in Figure~\ref{fig:NS_heating2}. These operators were chosen because they allow us to compare interactions with and without momentum or velocity suppressed scattering or annihilation cross sections. Specifically: 
\begin{itemize}
\item D1 (scalar) --  unsuppressed scattering cross section; $p$-wave suppressed annihilation.
\item D4 (pseudoscalar) -- $q_{\rm tr}^4$ suppressed scattering cross section; unsuppressed $s$-wave annihilation.
\item D5 (vector) -- unsuppressed scattering and annihilation cross sections.
\end{itemize}

Comparing the projected NS heating sensitivity with limits from terrestrial direct detection experiments (shown as green and orange curves in Fig.~\ref{fig:NS_heating}) we find similar behaviour for D1 and D5. This is expected, as both of these operators give rise to unsuppressed spin-independent DM-nucleon scattering cross sections. The $p$-wave suppression of the D1 annihilation cross section increases $\teq$ with respect to that for D5; nonetheless, equilibrium is reached relatively quickly compared to $t_\star$.  The D4 (pseudoscalar) interaction has dismal prospects of being observed in direct detection experiments, due to the severe  $q_{\rm tr}^4$ suppression for the scattering of non-relativistic DM. In contrast, NS heating has much greater sensitivity.

Note that the time required for complete thermalisation (grey contours) is much longer for D4 (momentum suppressed scattering) than for D1 and D5. In fact, for operator D4, full thermalisation is not achieved for most of the interesting parameter space. This illustrates the importance of correctly identifying $\teq$ as the timescale on which full heating is achieved, rather than the much longer $\tth$.

\section{Conclusions}
\label{sec:conclusions}

We have studied the capture of dark matter (DM) in neutron stars (NSs), and its subsequent thermalization and annihilation, giving detailed calculations of the timescales involved. These include the time required for DM to deposit $99\%$ of its initial kinetic energy through scattering processes, $\tkin$; the time required for the DM to thermalize within the NS core, $\tth$; and the timescale on which capture-annihilation equilibrium is reached, $\teq$. 
Our key conclusions are:

\begin{itemize}
    \item
    {\it The majority of the DM kinetic energy is deposited very quickly, on a timescale much shorter than $t_{\rm therm}$.} We find that over 99\% of the kinetic energy is deposited in a matter of days for DM captured in the geometric limit.  
    
    \item 
    The DM thermalization time depends strongly on DM mass, the NS temperature, and the scattering cross section.  For a momentum-independent cross-section, thermalization will always occur within about a Myr, and much faster for either very light or heavy DM.  In the case of momentum-suppressed scattering cross-sections, full thermalization would take longer than the age of the Universe for most of the parameter space.
   
    \item
    {\it Capture-annihilation equilibrium, and hence maximal annihilation heating, can be achieved without full thermalization.} We have outlined how to modify the usual capture-annihilation criterion to handle the case of partially thermalized DM.    
    \item 
    {\it Capture-annihilation equilibrium can be achieved on a short timescale for all interaction types, even those for which the scattering or annihilation cross sections are suppressed when the DM is non-relativistic.} For scattering cross sections that saturate the capture rate, we find that capture-annihilation equilibrium is typically reached on a timescale of less than 1 year for vector interactions and $10^4$ years for scalar interactions. 
\end{itemize}
These findings imply that even though the DM may not fully thermalize in the star, it can nonetheless lead to appreciable heating.   
In particular,  even interactions for which the scattering cross section is momentum-suppressed in the non-relativistic limit, or the annihilation cross section is velocity suppressed, maximal heating can be achieved for a wide region of parameter space. This enables future NS observations to provide a potential probe of interactions for which conventional direct detection experiments are insensitive.


\section*{Acknowledgements}
NFB and GB were supported by the Australian Research Council through the ARC Centre of Excellence for Dark Matter Particle Physics, CE200100008. SR acknowledges partial support from the UK STFC grant ST/T000759/1 and from the Fermi National Accelerator Laboratory (Fermilab), a U.S.
Department of Energy, Office of Science, HEP User Facility.
MV was supported by an Australian Government Research Training Program Scholarship. 

\newpage
\appendix

\section{Interaction rate in the zero temperature approximation}
\label{sec:pauliblockingle}

In this section,  we calculate the interaction rate in the zero temperature approximation for $\Msq = \alpha \, t^n$, where $n=0,1,2$ and $\alpha$ is a constant, in the low energy, Pauli suppressed regime where $K_\chi=E_\chi-m_\chi<\kinFi$. 
We assume the simplest scenario of constant target mass and point-like targets, as justified in Section~\ref{sec:thermstandard}. 

In this approximation, the interaction rate in the energy regime relevant for thermalization is given by~\cite{Bell:2020jou}
\begin{equation}
\Gamma^{-}(K_\chi,\Tstar=0) \propto \, \frac{1}{2^7\pi^3 K_\chi k }  \int_0^{K_\chi} q_0 dq_0 \,\, \int \frac{t_E^n dt_E }{\sqrt{q_0^2+t_E}}\Theta(\kinFi-q_0),
\label{eq:gammafinal}
\end{equation}
where $t_E=-t$,   and 
\begin{equation}
\int \frac{t_E^n dt_E }{\sqrt{q_0^2+t_E}} = 2D_n(q_0^2,t_E)\sqrt{q_0^2+t_E}.
\end{equation}
The $D_n$  functions can be found in Appendix B of  ref.~\cite{Bell:2020jou}. 
As we shall see, the integration intervals in Eq.~\ref{eq:gammafinal} depend on whether or not Pauli blocking suppresses any part of the thermalization process.
In both cases, we can find simple analytic approximations to these integrals. 
The minimal DM mass for which Pauli blocking is never in effect is denoted by $\mcrit$. 

We first consider the case where $m_\chi \lesssim \mcrit$. 
For the cases of $\mu \ll \kinFi/K_\chi$ or $\mu \gg K_\chi/\kinFi$, 
$\Gamma^{-}$ at first order in $K_\chi$ is given by
\begin{eqnarray}
\Gamma^{-}(K_\chi) &\sim& \frac{\alpha}{2^7\sqrt{2}\pi^3m_\chi^{3/2}K_\chi^{1/2}}\int_0^{K_\chi}q_0 dq_0 \left(\int_{t_{E}^{-}}^{t_{E}^{+}} \frac{t_E^n dt_E }{\sqrt{q_0^2+t_E}} \right)\\
&=& \frac{\alpha}{2^6\sqrt{2}\pi^3m_\chi^{3/2}K_\chi^{1/2}}\int_0^{K_\chi}q_0 dq_0 \left(\sqrt{q_0^2+t_E^{+}}D_n(q_0^2,t_E^{+})-\sqrt{q_0^2+t_E^{-}}D_n(q_0^2,t_E^{-})\right), \label{eq:intratelowe}
\end{eqnarray}
where $t_E^\pm$ are defined in ref.~\cite{Bell:2020jou}. 
For matrix elements independent of $t$ ($n=0$), we have $D_0(q_0^2,t_E^{\pm})=1$  and this result simplifies to
\begin{eqnarray}
\Gamma_{n=0}^{-}(K_\chi) 
&\sim&  \frac{|\overline{M}|^2}{2^6\pi^3m_\chi}\int_{0}^{K_\chi} dq_0 q_0 \left[\sqrt{2\left(1+\sqrt{1-\frac{q_0}{K_\chi}}\right)-\frac{q_0}{K_\chi}}-\sqrt{2\left(1-\sqrt{1-\frac{q_0}{K_\chi}}\right)-\frac{q_0}{K_\chi}}\right] \nonumber \\
&=& \frac{|\overline{M}|^2}{120\pi^3m_\chi}K_\chi^2.
\end{eqnarray}
We can rewrite the previous expression in terms of the DM-baryon scattering cross-section using the following expression
\begin{equation}
\sigma_{i\chi}^{n=0} = \frac{\Msq}{16\pi m_i^2 (1+\mu)^2}, 
\label{eq:intraten0}
\end{equation}
giving the interaction rate at first order in $K_\chi$
\begin{equation}
\Gamma_{n=0}^{-}(K_\chi)  \sim \frac{2 m_i}{15}\frac{(1+\mu)^2}{\mu}K_\chi^2 \sigma_{i\chi}^{n=0}.
\end{equation}
This result has the same $K_\chi$ and $\mu$ scaling as that of ref.~\cite{Bertoni:2013bsa}.

Performing a similar analysis for $\Msq=\alpha (-t)^n$, $n=1,2$, 
we find 
\begin{equation}
\Gamma_{n=1}^{-}(K_\chi) 
\sim \frac{2\alpha }{105\pi^3}K_\chi^3,\qquad 
\Gamma_{n=2}^{-}(K_\chi) 
\sim \frac{4\alpha }{63\pi^3}m_\chi K_\chi^4. 
\end{equation}
The expressions for the cross sections for $n=1,2$ are 
\begin{equation}
\sigma^{n=1}_{i\chi} = \frac{\alpha}{16\pi m_i^2(1+\mu)^2}t_{max},\qquad
\sigma^{n=2}_{i\chi} = \frac{4}{3}\frac{\alpha}{16\pi m_i^2(1+\mu)^2}t_{max}^2.   
\end{equation}
These cross sections must be normalized to sensible momentum transfer. We take this reference point to be the surface of the star, such that 
\begin{equation}
      t_{max} \sim \frac{4m_\chi^2}{1+\mu^2}\frac{1-B(\Rstar)}{B(\Rstar)}.   
\end{equation}
The interaction rates for $n=1,2$ can then be written as
\begin{eqnarray}
    \Gamma_{n=1}^{-}(K_\chi) &\sim& \frac{8}{105 \pi^2} \frac{(1+\mu)^2(1+\mu^2)}{\mu^2} \sigma_{\rm surf} \, K_\chi^3 \frac{B(\Rstar)}{1-B(\Rstar)}, \label{eq:intraten1}\\
\Gamma_{n=2}^{-}(K_\chi) &\sim& \frac{1}{21 \pi^2} \frac{(1+\mu)^2(1+\mu^2)^2}{\mu^3} \frac{\sigma_{\rm surf}}{m_i} \, K_\chi^4\left[\frac{B(\Rstar)}{1-B(\Rstar)}\right]^2.
\label{eq:intraten2}
\end{eqnarray}

We now look at the interaction rate in the super-heavy DM mass regime, $m_\chi \gtrsim \mcrit$.
The exact value of $\mcrit$ will depend on the NS configuration. However, we can take some typical values relevant to thermalization to give an estimate of its value. Taking $K_\chi=10^3\K$, $\kinFi=200\MeV$, we see that
\begin{equation}
    m_\chi \ge \frac{2\kinFi(2m_i+\kinFi)}{ K_\chi} \sim \frac{4\kinFi m_i}{K_\chi} = m_\chi^{\rm crit}\sim 9.65\times10^9\GeV. 
\end{equation}
The maximum energy transfer in this regime will always be $\qomax<K_\chi$, with
\begin{equation}
    \qomax\sim K_\chi\left[2\sqrt{\frac{m_\chi^{\rm crit}}{m_\chi }} - \frac{m_\chi^{\rm crit}}{m_\chi } +\mathcal{O}\left(\left(\frac{m_\chi^{\rm crit}}{m_\chi}\right)^{\frac{3}{2}}\right)\right].
 \end{equation}
Performing a similar analysis as the $m_\chi \lesssim \mcrit$ regime leads to the following expression for $\Gamma^-$,
\begin{equation}
\Gamma^{-}(K_\chi) \sim \frac{|\overline{M}|^2}{2^7\sqrt{2}\pi^3m_\chi^{3/2}K_\chi^{1/2}}\int_0^{	\qomax}q_0 dq_0 \left(\int_{t_{E}^{-}}^{t_{\mu^-}^{+}} \frac{t_E^n dt_E }{\sqrt{q_0^2+t_E}} \right),  
\end{equation} 
where $t_{\mu^-}^+$ is defined in ref.~\cite{Bell:2020jou}. 
For the simplest case of constant $|\overline{M}|^2$ this results in 
\begin{align}
\Gamma^{-}_{n=0}(K_\chi) & \sim  
\frac{K_\chi \kinFi |\overline{M}|^2}{24\pi^3\mu^2m_i}\left[\sqrt{\frac{m_\chi^{\rm crit}}{m_\chi}}+\mathcal{O}\left(\frac{m_\chi^{\rm crit}}{m_\chi}\right)\right] \nonumber\\
& = \frac{|\overline{M}|^2(m_i\kinFi)^{3/2}}{12 \pi^3m_\chi^{5/2}}K_\chi^{1/2}.\label{eq:intraten0largem}
\end{align}

\section{Thermalization of super-heavy DM}
\label{sec:thermsuperheavy}

For DM that is heavier than the critical mass  $m_\chi\gtrsim m_\chi^{\rm crit}$,
the energy lost in each scatter is a tiny fraction of the total DM kinetic energy. Moreover, the average time between collisions is typically on the order of fractions of a second. This warrants the use of a continuous approximation in this regime rather than performing the discrete summation. The thermalization time is then found by integrating the rate at which the DM kinetic energy changes, 
\begin{equation}
    \frac{dK_\chi}{dt} = -\Gamma^{-}(K_\chi) \langle\Delta K_\chi\rangle,  
    \label{eq:contttherm}
\end{equation}
from the initial kinetic energy, $K_\chi=m_\chi\left(\frac{1}{\sqrt{B(r)}}-1\right)$, to the final value $T_{\rm eq}\ll m_\chi$. For a constant cross-section ($n=0$), we substitute  Eqs.~\ref{eq:intraten0largem} and \ref{eq:aveElossn0largem} into the expression above leading to
\begin{equation}
    \tthn{0} \sim \frac{9 \pi^2 m_\chi}{8 (\mbeff)^2 \kinFi^2 \sigma_{i\chi}^{n=0}}\log\left[\frac{m_\chi}{T_{\rm eq}}\left(\frac{1}{\sqrt{B(\Rstar)}}-1\right)\right].
    \label{eq:tthemheavy0}
\end{equation}
Taking the final temperature to be $T_{\rm eq}=10^3\K$ and $B(\Rstar)=0.5$, this yields 
\begin{equation}
    \tthn{0} \sim 1.7  \yrs \left(\frac{m_\chi }{10^{10}\GeV}\right)\left(\frac{0.5\;m_n}{\mbeff(0)}\right)^{2}\left(\frac{0.2\GeV}{\kinFi(0)}\right)^{2}\left(\frac{10^{-45}\cm^2}{\sigma_{i\chi}^{n=0}}\right).    
\end{equation}
Repeating for $d\sigma\propto t^n$ ($n=1,2$), we calculate the thermalization time for $n=1$ to be
\begin{eqnarray}
    \tthn{1} &\sim& \frac{9\pi^2 m_\chi}{ 64 \mbeff \kinFi^3 \sigma_{i\chi}^{n=1}}\left[\frac{1-B(\Rstar)}{B(\Rstar)}\right] \log\left[\frac{m_\chi}{T_{\rm eq}} \left(\frac{1}{\sqrt{B(\Rstar)}}-1\right)\right],\\
    &\sim& 3.5 \yrs\; \left(\frac{m_\chi}{10^{10} \GeV}\right)\left(\frac{0.5\;m_n}{\mbeff(0)}\right) \left(\frac{0.2\GeV}{\kinFi(0)}\right)^{3}\left(\frac{10^{-45}\cm^2}{\sigma_{i\chi}^{n=1}}\right),
\end{eqnarray}
and that for $n=2$ to be
\begin{eqnarray}
    \tthn{2} &\sim& \frac{5 \pi^2 m_\chi}{ 32 \kinFi^4\sigma_{i\chi}^{n=2}} \left[\frac{1-B(\Rstar)}{B(\Rstar)}\right]^2 \log\left[\frac{m_\chi}{T_{\rm eq}}\left(\frac{1}{\sqrt{B(\Rstar)}}-1\right)\right],\\
    &\sim& 3.5 \yrs \left( \frac{m_\chi}{10^{10} \GeV}\right) \left(\frac{0.2\GeV}{\kinFi(0)}\right)^{4}\left(\frac{10^{-45}\cm^2}{\sigma_{i\chi}^{n=2}}\right). 
\end{eqnarray}

\section{Thermalization time for $s$- and $t$-dependent interactions}
\label{sec:sdeptherm}

In Section~\ref{sec:thermstandard}, we assumed $\Msq\propto t^n$ when deriving analytical approximations for the thermalization timescale. To understand the behavior of the thermalization time for the operators in Table.~\ref{tab:operatorshe}, we can make use of the results for $t^n$ dependent interactions. For cross sections that are linear combinations of different powers of $t$, we can approximate the thermalization time using the previous results in the following way
\begin{eqnarray}
\Msq &=& a_0 + a_1 t + a_2 t^2,\\
\sigma &=& a_0\sigma_0 + a_1 \sigma_1 + a_2 \sigma_2,\\
\frac{1}{\tth} &\sim& \frac{a_0}{ \tthn{0}(\sigma_{i\chi}=\sigma_0)} + \frac{a_1} {\tthn{1}(\sigma_{i\chi}=\sigma_1)} 
 + \frac{a_2}{  \tthn{2}(\sigma_{i\chi}=\sigma_2)}. 
\label{eq:ttherm_weighted}
\end{eqnarray}
Hence, the inverse of the thermalization time will be given by a weighted linear combination of the inverse times for each contribution. As higher powers of $t$ require significantly longer thermalization times, for coefficients of similar size, the resulting sum will be dominated by the lowest power of $t$ appearing in  $\Msq$.  We can thus identify the dominant terms for operators D1-D4  based on power counting, which we have listed in Table~\ref{tab:operatorshe}.

 For $s$-dependent amplitudes, we can in principle use the interaction rates calculated  in Appendix A of ref.~\cite{Bell:2020lmm}, perform a series expansion in $K_\chi$  and repeat the same procedure outlined in Section~\ref{sec:thermstandard} for $s$-independent matrix elements. Interestingly, we find that for the purpose of calculating the thermalization time, there is an easier way to obtain the correct result. One can indeed check that, at zero order in $\kinFi/\mbeff$, the resulting time for $s^1, s^2$ is equivalent to the constant case, with the matrix element calculated by setting 
\begin{equation}
    s\rightarrow (m_\chi+\mbeff)^2,
    \label{eq:ssubst}
\end{equation}
while the $s t $ case has a result equivalent to the $t$ case, with the matrix element calculated using the same substitution. This is, in practice, equivalent to setting both the DM and neutron targets at rest. There is, however, an important exception, when it comes to calculating the thermalization time of a linear combination of these terms. In particular, when the amplitudes at order $\mathcal{O}(t^0)$, are proportional to combinations of $1,s,s^2$ such as
\begin{eqnarray}
s-(m_\chi+\mbeff)^2,\nonumber\\
\left[s-(m_\chi+\mbeff)^2\right]^2,\nonumber\\
\left[s-(\mbeff)^2-m_\chi^2\right]^2-4(\mbeff)^2 m_\chi^2.\label{eq:m2veldep}
\end{eqnarray}
All these combinations give a null result after applying substitution \ref{eq:ssubst}. In such a case, one may think that the dominant term is given by some remaining $t^n$ term. It is worth noting that the expressions in  Eq.~\ref{eq:m2veldep}  appear in operators that, at low energy, are known as  velocity-dependent, because their matrix elements are proportional to positive even powers of the DM-target relative speed. Consequently, it is important not to neglect the motion of the targets in the neutron star, moving at relativistic speeds that are of the order of the Fermi velocity $v_F^2=2\kinFi/\mbeff$. In those cases, one should instead set $s$ to\footnote{We assume that $\mu\gg \mbeff/m_\chi^{\rm crit}$ when making this substitution.}
\begin{equation}
    s\rightarrow (m_\chi+\mbeff)^2+2m_\chi\kinFi.
    \label{eq:ssubstmu}
\end{equation}

In summary, operators D5, D8 and D9 can be safely expanded using \ref{eq:ssubst}, while operators D6, D7 and D10 have velocity-dependent amplitudes and require Eq.~\ref{eq:ssubstmu}. 
The dominant terms for each operator can be found in Table~\ref{tab:operatorshe}. 
For equal values of the leading term in $\Msq$, the thermalization time for each operator will be the same as the relevant $t^n$ power law.

\section{Temperature distribution of captured dark matter}
\label{sec:minTempDerivation}

As seen in Fig.~\ref{fig:thermtime}, interactions that depend on the momentum transfer, namely $d\sigma \propto t^n$ with $n = 1,2$, there are regions of the DM mass parameter space where thermalization does not occur within the age of the star. For the DM masses and NS temperatures of interest, this region of non-thermalization always occurs in the $m_\chi\ll \mcrit$ regime.
From Eqs.~\ref{eq:thermtimen0}, \ref{eq:thermtimen1} and \ref{eq:thermtimen2},  we can estimate the time required for the DM to reach a kinetic energy $K_\chi$ 
\begin{equation}
    t_{K_\chi} \propto \frac{1}{K_\chi^{n+2}}.
\end{equation}
If the DM does not thermalize within the age of the star, it will instead reach a minimum temperature, $K_\chi^{\mathrm{min}}$.  Comparing the time required to achieve this temperature to the thermalization time, $\tth$ i.e. to have reached the equilibrium temperature $\Teq$, we find 
\begin{equation}
    \frac{t_{K_\chi^\mathrm{min}}}{\tth}  \sim \left( \frac{\Teq}{K^{\rm min}_\chi} \right)^{n+2}. 
\end{equation}
Accounting for the case where the DM reaches thermalization, we can write $K_\chi^\mathrm{min}$
\begin{align}
    K^{\rm min}_\chi & \sim \Teq \left(\max\left[ 1,\frac{\tth}{t_{K_\chi^\mathrm{min}}}\right ]\right)^{\frac{1}{n + 2}}\\
           & \approx \Teq \left( 1 + \frac{\tth}{t_{K_\chi^\mathrm{min}}}\right )^{\frac{1}{n + 2}}. 
\end{align}
The population of captured DM will have a distribution of energies at any given time, with this distribution being peaked at this minimum energy.
As the orbital periods of the DM will be much shorter than the average time between interactions, the DM will be able to virialize between each interaction. Therefore, we can treat the DM as being contained within an isothermal sphere with temperature $K^{\rm min}_\chi > \Teq$. 

Finally, it is worth noting that at times  $t>\tth$, even though the thermalization condition has been reached, the captured DM would consist of two components: a fraction of it (whose amount depends on time) would be in thermal equilibrium with the NS at temperature $T_{\rm eq}$; and another component still in the cooling down process. Assuming a  capture rate constant over time, the fraction of thermalized DM is 
\begin{equation}
    f_{\rm therm}(t) = \frac{t-t_{\rm therm}}{t}.
\end{equation}

\section{Quark-level vs hadron-level annihilation cross sections}
\label{sec:quarkhadron}

The annihilation cross sections shown in Table~\ref{tab:annCS} are for DM annihilation to quark final states.  More properly, we should consider the hadron-level annihilation cross section.
However, we are primarily concerned with the capture-annihilation equilibrium timescale, and not the details of the annihilation process. Therefore, if the annihilation rate to hadrons is not significantly different from the quark level result, this subtlety can be avoided. 

To check the validity of the quark-level approximation, we estimate the annihilation rate to hadrons, working at lowest order in Chiral Perturbation Theory. We use couplings to the meson octet obtained from  ref.~\cite{Kumar:2018heq}; for annihilation to baryons, the operators listed in Table~\ref{tab:operatorshe} are used. 
For DM masses in the range $m_\pi < m_\chi \lesssim m_{\rm charm} = 1.27\GeV$, we find that the cross section for annihilation to hadrons differs by less than an order of magnitude than that for annihilation to quarks. For larger DM mass, the difference is negligible. Therefore, to simplify the discussion, we consider DM annihilation to quark final states for DM masses above the pion mass.

Below the pion mass, the only kinematically allowed DM annihilation channels would be to leptons or photons. The size of the DM couplings to these states would, in general, be unrelated to the DM-quark couplings we have assumed. (They are expected to be non-zero, because they would be induced at loop level~\cite{Bell:2019pyc}, even if absent at tree level.)
However, due to the considerable Fermi energies of the electrons and muons near the centre of the NS, these channels will be Pauli blocked for the whole DM mass range below $m_\pi$, forbidding these annihilations from occurring. To remain as model-independent as possible, we will not consider lepton and photon annihilation channels.

\begin{figure*}
    \centering
    \includegraphics[width = 0.496\textwidth]{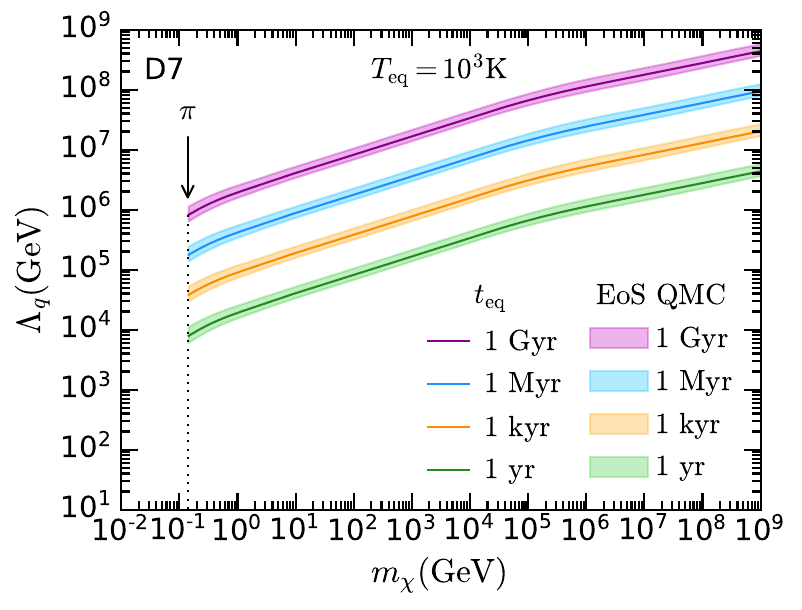}
    \includegraphics[width = 0.496\textwidth]{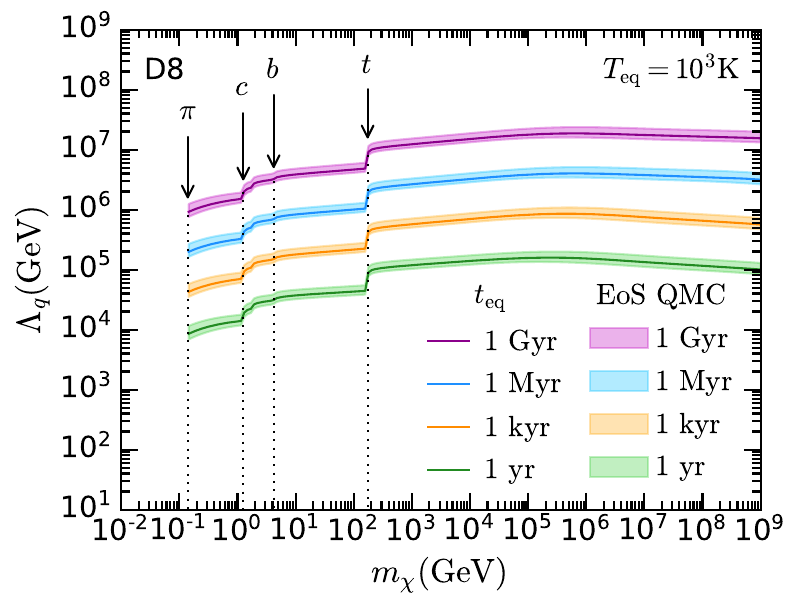}    
    \caption{
    Contours of the capture annihilation timescale, $t_{\rm eq}$,  in the $\Lambda_q-m_\chi$ plane for operators D7 (left) and D8 (right) and $\Tstareq=1000\K$. 
    Solid lines represent the calculation for the NS benchmark model QMC-2, and shaded regions denote the variation with the NS choice for the QMC EoS family. 
    Dotted lines in the right panel indicate the mass thresholds for various annihilation channels.  
    }
    \label{fig:ann_xs_plots}
\end{figure*}

Figure~\ref{fig:ann_xs_plots} shows $t_{\rm eq}$ contour lines in the $\Lambda_q-m_\chi$ plane for the NS benchmark model QMC-2, $\Tstareq=1000\K$ and two representative operators D7 (left) and D8 (right). 
Operators whose thermally averaged annihilation cross section $\sigmav$ has a  $m_q/m_\chi$ leading order term, namely D1-D4 and D8 (see Table~\ref{tab:annCS}), exhibit a sudden change in the slope wherever a new annihilation channel opens (see dotted lines on the right panel of Fig.~\ref{fig:ann_xs_plots}). Note that the higher the cutoff scale $\Lambda_q$, the lower the scattering and annihilation cross sections, resulting in a larger $t_{\rm eq}$ timescale. For lower $\Teq$ temperatures, DM requires more time to reach both equilibrium conditions, thermalization and capture-annihilation. The variation of these results with respect to the NS configuration amounts at most to a factor of $\sim2$ in the $t_{\rm eq}$ contours (see shaded regions in the left panels) from the lightest configuration (QMC-1, $1\Msun$) to the heaviest (QMC-3, $1.9\Msun$) for most operators, with the sole exception of D4 for which this factor rises up to $\sim 2.4$. 

\section{Capture-annihilation equilibrium for the EFT operators}
\label{sec:resultsEFTop}

\begin{figure*}
    \centering
    \includegraphics[width=\textwidth]{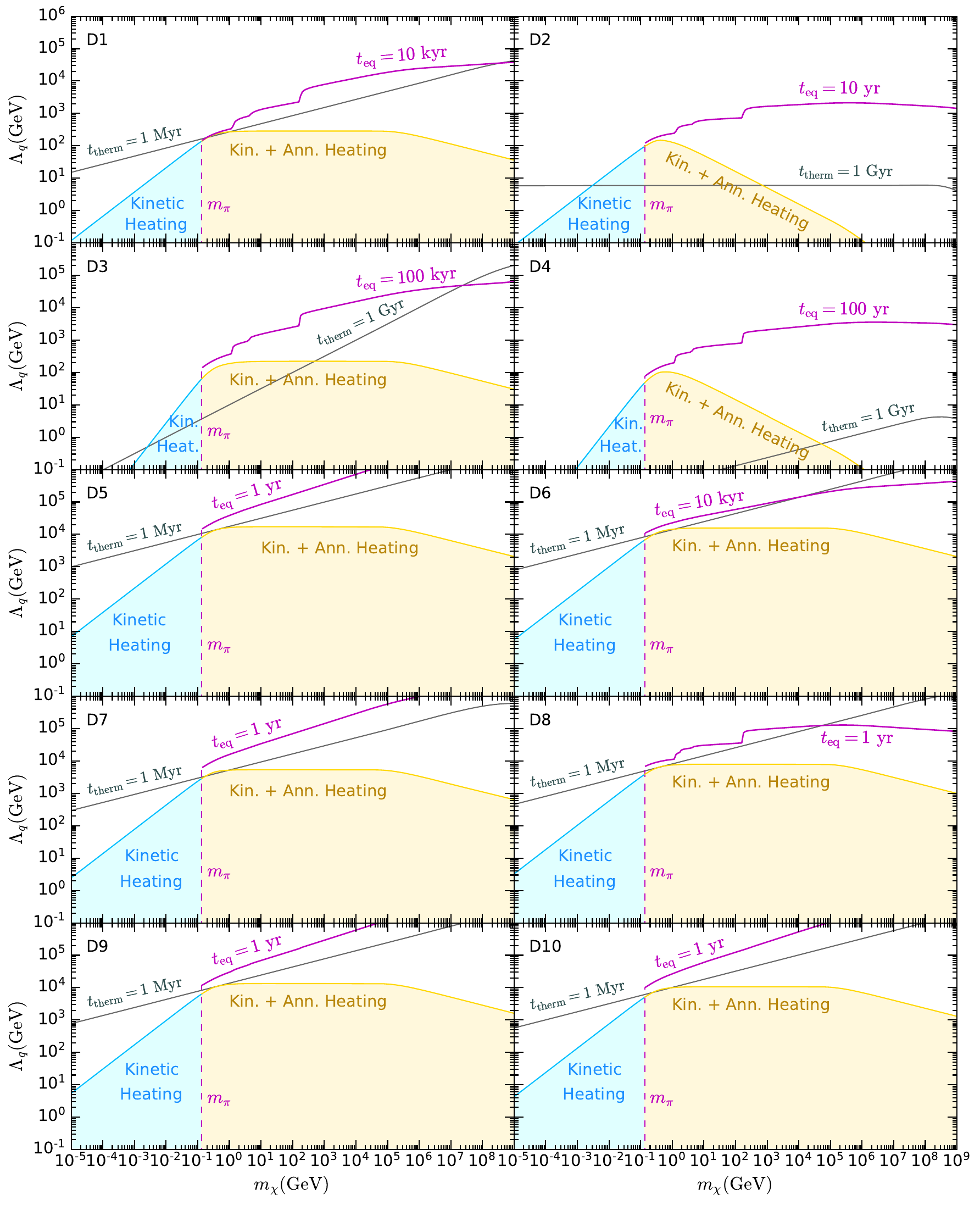}    
    \caption{Projected NS heating sensitivity for maximal capture efficiency, for the full set of DM-baryon interactions described by the EFT operators of Table~\ref{tab:operatorshe}. We have used the QMC-2 ($1.5\Msun$) benchmark NS configuration.  
    Color coding as in Fig.~\ref{fig:NS_heating}. 
    }
    \label{fig:NS_heating2}
\end{figure*}

In Fig.~\ref{fig:NS_heating2}, we show isocontours of maximal capture (yellow and light blue lines) and  capture-annihilation equilibrium (magenta lines) timescale in the $\Lambda_q-m_\chi$ plane for all  EFT operators.  
Values of $\Lambda_q$ below the $t_{\rm eq}$ lines result in smaller capture-annihilation equilibrium timescales. 
We can see in Fig.~\ref{fig:NS_heating} that for all operators the $t_{\rm eq}$ timescale is always smaller than the time required for captured DM to thermalize. Captured DM achieves the steady state condition in a timescale as short as $\sim1\yr$ (D1, D6-D10) or as long as $10^5\yr$ (D3). 
Note  that the displayed values of $t_{\rm eq}$ are the  values for which the entire parameter space relevant for capture reach equilibrium with annihilation. 

For D1 and D6-D10, we observe that captured DM achieves thermal  equilibrium in less than $\sim 1\Myr$ (grey lines). 
On the other hand, as expected from Fig.~\ref{fig:thermtime} captured DM whose interactions are momentum suppressed, namely operators D2-D4, would never thermalize to the temperature expected from DM induced heating in $1\Gyr$, or even in less than the age of the 
Universe,
with the sole exception of the corner region of very light DM $m_\chi\lesssim 2\MeV$ and an even narrower corner of the parameter space for D4. 

Therefore, for all EFT operators,  the energy released in the annihilation process adds up to the energy deposited via capture increasing the DM induced heating  
for $m_\chi\gtrsim m_\pi$ (yellow shaded area). 
Recall that we have made no assumptions about the energy scale that controls DM interactions with leptons. 
For DM of mass below $m_\pi$ (light blue lines), at least kinetic heating is expected to contribute to the star luminosity.

\label{Bibliography}

\lhead{\emph{Bibliography}} 

\bibliography{Bibliography} 

\end{document}